\documentclass[aps,prd,reprint,onecolumn,notitlepage,superscriptaddress]{revtex4-1}

\usepackage{lipsum}
\usepackage{amsmath}
\usepackage{graphicx}
\usepackage{color}
\usepackage{tikz}
\usepackage{verbatim}
\usepackage{hyperref}
\usepackage{amssymb}
\usepackage{bm}
\newcommand{\rom}[1]{\textup{\uppercase\expandafter{\romannumeral#1}}}

\allowdisplaybreaks

\begin{document}

\title{Ghost Free Theory in Unitary Gauge: A New Candidate}
\author{Pawan Joshi\footnote{email:pawanjoshi697@iiserb.ac.in}
       }

\affiliation{Indian Institute of Science Education and Research Bhopal,\\ Bhopal, Madhya Pradesh - 462066, India}

\author{Sukanta Panda \footnote{email:sukanta@iiserb.ac.in}
       }

\affiliation{Indian Institute of Science Education and Research Bhopal,\\ Bhopal, Madhya Pradesh - 462066, India}

\author{Archit Vidyarthi \footnote{email:archit17@iiserb.ac.in}
       }

\affiliation{Indian Institute of Science Education and Research Bhopal,\\ Bhopal, Madhya Pradesh - 462066, India}

\begin{abstract}
We propose an algebraic analysis using a 3+1 decomposition to identify conditions for a clever cancellation of the higher derivatives, which plagued the theory with Ostrogradsky ghosts, by exploiting some existing degeneracy in the Lagrangian. We obtain these conditions as linear equations (in terms of coefficients of the higher derivative terms) and demand that they vanish, such that the existence of nontrivial solutions implies that the theory is degenerate. We find that, for the theory under consideration, no such solutions exist for a general inhomogeneous scalar field, but that the theory is degenerate in the unitary gauge. We, then, find modified FLRW equations and narrow down conditions for which there could exist a de Sitter inflationary epoch. We further find constraints on the coefficients of the remaining higher-derivative interaction terms, based on power-counting renormalizability and tree-level unitarity up to the Planck scale.

\end{abstract}

\maketitle
\section{Introduction}
The scalar-tensor theories (STTs) are one of several alternatives to General Relativity (GR), that have garnered special interest in their ability to provide a unified mathematical description for modified gravity theories. STTs introduce an additional scalar degree of freedom (DOF) along with the two existing tensor DsOF present in GR \cite{PhysRevD.37.3406, Peebles:1987ek, Carroll:1998zi}. This extra degree of freedom can address numerous problems, such as dark energy, dark matter, inflation, big-bang singularity, cosmic magnetic field, etc, for which GR proves insufficient.

Usually, STTs don't provide a healthy DOF, and the new scalar can generally be associated with instability. When higher derivatives appear in the non-degenerate Lagrangian, the associated instability is called the Ostrogradsky ghost \cite{Ostrogradsky:1850fid}. Mostly these theories include terms involving a curvature tensor in the linear order coupled with higher derivatives of the scalar field. The instability can be identified either in the Lagrangian with higher-order derivatives, or through the associated Hamiltonian with the linear momenta \cite{Woodard:2015zca, Chen:2012au, Joshi:2019cyk, Joshi:2022dpz}. However, there exist stable theories with a Lagrangian containing higher derivatives where the ghosts are avoided by ensuring that the higher derivatives are not a part of the equations of motion. When a degeneracy is present in the Lagrangian, it falls under the classification of DHOST (degenerate higher-order derivative scalar tensor) theories. Common examples include Horndeski theories \cite{Horndeski:1974wa, Nicolis:2008in,Deffayet:2011gz, Kobayashi:2011nu}, beyond Horndeski \cite{Gleyzes:2014dya,Gleyzes:2014qga}, and others such as \cite{Langlois:2015cwa, Motohashi:2016ftl, BenAchour:2016fzp,Chagoya:2018lmv} (for a review, refer to\cite{Langlois:2018dxi,Kobayashi:2019hrl}). The DHOST action has been studied in the context of gravity and cosmology in \cite{Langlois:2015skt, Nicolis:2008in, Deffayet:2011gz, Kobayashi:2011nu, Gleyzes:2014dya, Gleyzes:2014qga, Langlois:2015cwa, Motohashi:2016ftl, Koyama:2015vza, Ilyas:2020qja, Motohashi:2019sen}.
Besides these, the theories with higher-order curvature are studied in Chern-Simons gravity \cite{Lue:1998mq, Jackiw:2003pm}, ghost-free parity violation theory \cite{Deruelle:2012xv}, and higher derivative metric theories\cite{Crisostomi:2017ugk}).

Another class of such higher derivative theories is the UDHOST theories, wherein we specifically work in the unitary gauge. The unitary gauge is a choice for which the scalar field is taken to be spatially uniform, i.e., the spatial derivative of the scalar field vanishes in 3+1 decomposition. This assumption dramatically reduces the effort of checking complex and tidal calculations for degeneracy conditions by simplifying the structure of the Lagrangian. Recently, we put forward a ghost-free theory in unitary gauge \cite{Joshi:2021azw} which contains, up to quadratic order, the most general combination of curvature couplings involving, up to second order, derivatives of a scalar field.

 UDHOST defines a broader class of degenerate theories than DHOST \cite{DeFelice:2018ewo, DeFelice:2021hps, DeFelice:2022xvq}. Also, since the unitary gauge appears naturally in the standard cosmological background (homogeneous space-time) by assuming that the scalar field has only a time-like gradient, it exhibits suitable features in the context of inflation \cite{Creminelli:2006xe, Cheung:2007st}. Recently, effective field theory (EFT) of perturbations on an arbitrary background metric with a time-like scalar profile has been studied in \cite{Mukohyama:2022enj, Mukohyama:2022enj, Mukohyama:2022skk}. Another theory studied under the unitary gauge is the spatially covariant theory of gravity (SCG)\cite{Gao:2014soa, Gao:2014fra, Fujita:2015ymn, Gao:2018znj, Gao:2019lpz, Gao:2018izs, Gao:2019liu, Gao:2020yzr, Gao:2020qxy, Hu:2021bbo}. Some examples of SCGs include Horava Lifshitz gravity\cite{Horava:2009uw, Blas:2009qj}, Cuscuton theory\cite{Afshordi:2006ad,Afshordi:2007yx}, and its generalization\cite{Gomes:2017tzd,Iyonaga:2018vnu}. Recently, in \cite{DeFelice:2018ewo, DeFelice:2021hps, PhysRevLett.129.031103}, it was shown that in a general coordinate system, the extra mode appearing in UDHOST theories is non-propagating, called a shadowy mode.


In this work, we modify the Lagrangian $L_{1-6}$ of \cite{Joshi:2021azw}. This modified Lagrangian has two additional first derivatives of the scalar field and the rest of the term of $L_{1-6}$. We are looking for any degeneracies present in this combination, for which, we decompose the modified Lagrangian using the 3+1 formalism. We obtain different relations (linear equations) among Lagrangian coefficients to remove the higher derivative terms from the equation of motion. After setting up a particular connection between their coefficients, we show that our theory has no instability associated with higher-derivatives in unitary gauge. We, then, analyze the physical properties of the theory, in the context of inflation, and provide suitable constraints, based on tree-level unitarity arguments, on the parameters introduced with these higher-derivative modifications.

The paper is organized as follows: In section \ref{3+1}, 3+1 decomposition or the Arnowitt-Deser-Misner (ADM) formalism is reviewed; Section \ref{degen} addresses important concepts of \cite{Joshi:2021azw}, and introduces the modified Lagrangian which is further analyzed in section \ref{l16fd}; In section \ref{unitary}, we analyze our theory in unitary gauge and discuss the ghost-free case; In section \ref{frwback}, we analyze the same Lagrangian for FLRW background, before proceeding to find conditions necessary to ensure a de Sitter inflationary phase in section \ref{dsinflation}; Then, we find the correction to the 2-scalar$\to$2-scalar tree-level scattering amplitude in section \ref{unitarity}, where we further impose constraints on coupling parameters using unitarity and power-counting renormalizability arguments as constraints; Finally, we summarise our results in section \ref{concl}.

\section{3+1 decomposition}\label{3+1}
Let us consider a four-dimensional space-time $\mathcal{M}$ with a metric $ g_{a b} $. This space-time can be split into three non-intersecting spacelike surfaces, characterized by the spatial 3-dimensional metric $h_{a b}$. These 3-surfaces are linked by a  normal vector $ n^{a} $, which satisfies the normalization condition $n_{a}n^{a}=-1$. By using this formalism, the 4-dimensional metric can be expressed using $ h_{a b}$ and $ n^{a} $ (For more details, refer to \cite{Baumgarte:2010ndz}),
\begin{equation}
g_{a b}= h_{a b}- n_{a} n_{b}.\label{metrich}
\end{equation} 
The induced metric and the normal vector can be used to decompose any covariant tensor fields. For example, any vector $A_{a}$, can be decomposed as,
\begin{eqnarray}
A_{a}=\mathcal{A}_{a}- A_{*}n_{a},
\end{eqnarray}
where $\mathcal{A}_{a}$ and $A_{*}$ are the purely spatial $(n^{a}\mathcal{A}_{a}=0)$ and one-normal part of the vector $A_{a}$, respectively. These are mathematically expressed as follows,
\begin{eqnarray}
\mathcal{A}_{a}=h_{a}^{b} A_{b}, \qquad\qquad A_{*}= A_{a}n^{a}.
\end{eqnarray}
Now, another important quantity is the time direction vector $t^{a} =\frac{\partial}{\partial t}$, which connects the points that have the same spatial coordinates on neighboring slices. Its form can be given as,
\begin{equation}
    t^{a}=Nn^{a}+N^{a},
\end{equation}
where $ N^{a} $ is the shift vector, and $N$ is the lapse function. 
 Since $A_{a}=\nabla_{a}\Phi$, so using the property, $\nabla_{a}A_{b} =\nabla_{b} A_{a}$,  the 3+1 decomposed form of  the $\nabla_{a}A_{b}$ is given as,
\begin{equation}
    \nabla_{a} A_{b} =\mathcal{D}_{a} \mathcal{A}_{b} - A_{*} K_{a b} +n_{a}(K_{b c} \mathcal{A}^{c} -\mathcal{D}_{b} A_{*} )+ n_{b}(K_{a c} \mathcal{A}^{c} -\mathcal{D}_{a} A_{*} ) +n_{a} n_{b}(\mathcal{L}_{n}{A}_{*}- \mathcal{A}_{c} a^{c}),\label{main} 
\end{equation}
where $\mathcal{D}_{a}$ denotes a spatial derivative, $a_{c}=n^{a}\nabla_{a}n_{c}$ is the acceleration vector, and $ K_{a b}$ is extrinsic curvature tensor. Its form is,
\begin{eqnarray} 
K_{a b}=\frac{1}{2N}\left( \dot{h_{a b}}-\mathcal{D}_{a} N_{b}-\mathcal{D}_{b}N_{a}\right), 
\end{eqnarray}
and the form of 
 $\mathcal{L}_{n}{A}_{*}$ is,
\begin{eqnarray}
 \mathcal{L}_{n}{A}_{*}=\frac{1}{N}(\dot{A_{*}}-N^{c}\mathcal{D}_{c}A_{*}).
\end{eqnarray}
where a `$\Dot{\ }$' above the variable indicates a time derivative. $\dot{h}_{a b}$ is the first-order time derivative of the metric, and $\dot{A}_{*}$ involves the second-order time derivative of the scalar field.

There are a lot of variables in (ref{main}]), so here we introduce a new set of variables \{$ U_{a b}$, $ Y_{b}$, $ Z_{*}$\} to simplify it, where $ U_{a b}=(\mathcal{D}_{a} \mathcal{A}_{b} - A_{*} K_{a b}) $ behaves as a symmetric spatial tensor, $ Y_{b}=(K_{b c} \mathcal{A}^{c} -\mathcal{D}_{b} A_{*}$) transforms like a spatial vector, and $ Z_{*}=(\mathcal{L}_{n}{A}_{*}- \mathcal{A}_{c} a^{c})$ contains a second order time derivative of the scalar field. Therefore, in terms of these new variables, (\ref{main}) becomes,
\begin{eqnarray}
(\nabla_{a} A_{b})= \dfrac{1}{N}  n_{a} n_{b} Z_{*}-U_{ab}-n_{a}Y_{b} \ -n_{b} Y_{a}.
\label{R1}
\end{eqnarray}
 We will use a similar methodology for separating higher derivatives with first-order and spatial derivatives of the curvature tensors. For this, we write the 3+1 decomposed form of the curvature tensors. First, the Riemann tensor is expressed as follows,
\begin{eqnarray}
\begin{aligned}
\mathfrak{R}_{a b c d} =& \mathcal{R}_{a b c d}+K_{a c} K_{b d}- K_{a d}K_{b c}, \\
V_{a b c}   =& \mathcal{D}_{a}K_{b c}-\mathcal{D}_{b}K_{a c}, \label{sdfe}\end{aligned}
\end{eqnarray}
where $\mathcal{R}_{abcd}$ is an intrinsic Riemann curvature tensor. The spatial ($\mathfrak{R}_{a b c d} $) and one-normal ($V_{a b c} $) decomposed parts are known as Gauss and Codazzi relations, respectively.

The two-normal $({}_{\perp} R_{a n b n})$ decomposition of Riemann tensor is given as,
\begin{eqnarray}
\begin{aligned}
{}_{\perp} R_{a n b n}=&  K_{a u} K^{u}_{b}- \mathcal{L}_{n}  K_{a b} + \mathcal{D}_{(a} a_{b)}+a_{a} a_{b},\label{dfd}
\end{aligned}
\end{eqnarray}
where $ \mathcal{L}_{n}K_{a b}$ is the Lie derivative of  extrinsic curvature tensor,
\begin{eqnarray}
\mathcal{L}_{n}K_{a b}=\frac{1}{N}(\dot{ K}_{a b}-\mathcal{L}_{\Vec{N}}K_{a b}).
\end{eqnarray}
Where $\mathcal{L}_{\Vec{N}}$ is the Lie derivative w.r.t the shift vector,  $\dot{ K}_{a b}$ contains second order time derivative of the induced metric $h_{ab}$.
 We  now introduce $\mathcal{F}_{ab}$,
  \begin{equation}
   \mathcal{F}_{ab}= K_{a u} K^{u}_{b} + \mathcal{D}_{(a} a_{b)}+a_{a} a_{b},
\end{equation} 
takes care of the first derivative of the metric and spatial derivative part. Thus, (\ref{dfd}) becomes,
\begin{eqnarray}
\begin{aligned}
    {}_{\perp} R_{a n b n}= \mathcal{F}_{ab}- \mathcal{L}_{n}  K_{a b}.\label{R6}
\end{aligned}
\end{eqnarray}
  To obtain the 3+1 decomposed form of the Ricci tensor, $ R_{a b}=g^{cd}R_{ac bd}$,  one can use the aforementioned relation of the Riemann tensor.
The purely spatial part of the decomposed Ricci tensor is,
\begin{eqnarray} 
\begin{aligned}
{}_{\perp}R_{ab}&= \mathcal{R}_{a b}+ K_{a b} K - 2K_{a s} K^{s}_{b}+  \mathcal{L}_{n}  K_{a b} -  \mathcal{D}_{(a} a_{b)}- a_{a} a_{b}\\&=  \mathcal{L}_{n}  K_{a b}+F_{a b}, \label{ps}
\end{aligned}
\end{eqnarray}
where $F_{a b}$ contains  first order derivatives of $h_{ab}$ and purely spatial terms,
\begin{equation}
  F_{a b} =  \mathcal{R}_{a b}+ K_{a b} K - 2K_{a s} K^{s}_{b}-  \mathcal{D}_{(a} a_{b)}- 2a_{a} a_{b}.
\end{equation} 
Next, the one-normal projection of Ricci tensor denoted by a vector notation $V_{b}$ is,
\begin{eqnarray}
{}_{\perp}R_{b n}= \mathcal{D}_{s}K_{b}^{s}- \mathcal{D}_{b} K = V_{b}. \label{on}
\end{eqnarray}
Finally, two-normal projections of the Ricci tensor can be expressed as,
\begin{eqnarray}
\begin{aligned}
{}_{\perp}R_{n n}&=  K_{s t} K^{s t}- h^{s t} \mathcal{L}_{n}  K_{s t} +  \mathcal{D}_{s} a^{s}+a_{s} a^{s}\\
&=\mathcal{F}_{2}-h^{s t} \mathcal{L}_{n}  K_{s t}. \label{tn}
\end{aligned}
\end{eqnarray}
Similarly, $\mathcal{F}_{2}$ is introduced to separate notation for showing first-order derivatives of $h_{ab}$ and purely spatial terms,
\begin{equation}
  \mathcal{F}_{2}=   K_{s t} K^{s t} +  \mathcal{D}_{s} a^{s}+a_{s} a^{s}.
\end{equation}
Now the full expression of Ricci tensor by combining eq.(\ref{ps}), eq.(\ref{on}) and eq.(\ref{tn}) becomes,
 \begin{eqnarray}
  R_{a b}=  \mathcal{L}_{n}  K_{a b}+F_{a b}
   -2 n_({a}V_{b)}+n_{a}n_{b}(\mathcal{F}_{2} -h^{s t} \mathcal{L}_{n}  K_{s t}), \label{R3}
\end{eqnarray} 
and the 3+1 decomposition of Ricci scalar $R=g^{ab}R_{ab}$  takes the form,
\begin{equation}
R= \mathcal{R}+ K^{2}-3 K_{a b} K^{a b}+ 2 h^{a b} \mathcal{L}_{n}  K_{a b} -2 \mathcal{D}_{b} a^{b}-2a_{b} a^{b}. \label{rs}
\end{equation}
Further, we introduce a scalar $\mathcal{F}_{1}$,
\begin{equation}
\mathcal{F}_{1}=\mathcal{R}+ K^{2}-3 K_{a b} K^{a b} -2  \mathcal{D}_{b} a^{b}-2a_{b} a^{b},
\end{equation}
which takes care of the first-order derivative of spatial metric $h_{ab}$ and purely spatial terms, using which eq.(\ref{rs}) can be simplified to,
\begin{equation}
R=  2 h^{a b} \mathcal{L}_{n}  K_{a b} +\mathcal{F}_{1}.\label{R2} \,
\end{equation} 
This form and the previously derived compact form of decomposed relations are beneficial for identifying higher-order derivative terms of the metric and scalar field. In the next section, we shall apply these relations.

\section{Degenerate Theory in Unitary Gauge}\label{degen}
In this section, we discuss the conditions needed to ensure the absence of ghosts in the higher derivative theory presented in \cite{Joshi:2021azw} by using the 3+1 decomposition.

\subsection{The Action}
Let us consider the action,
 \begin{eqnarray}
 S= \int d^{4}x \sqrt{-g} \  \tilde{H}^{\mu \nu \rho \sigma \alpha \beta \gamma \delta} R_{\alpha\beta\gamma\delta} \nabla_{\mu} \nabla_{\nu} \phi \,\nabla_{\rho} \nabla_{\sigma} \phi  \label{S0},
\end{eqnarray} 
  where, the tensor $ \tilde{H}^{\mu \nu \rho \sigma \alpha \beta \gamma \delta}$ is given as,
  \begin{eqnarray}
\tilde{H}^{\mu \nu \rho \sigma\alpha \beta \gamma \delta}= (D_{1} g^{\mu \rho} g^{\nu \sigma} +D_{2} g^{\mu \sigma} g^{\nu \rho})  g^{\alpha\gamma} g^{\beta\delta}  + (D_{3} g^{ \beta \rho} g^{ \mu \nu } g^{\delta \sigma}  +D_{4}  g^{\mu \beta} g^{\delta \rho} g^{\nu \sigma})g^{\alpha\gamma} \nonumber \\ + D_{6} g^{  \mu \alpha} g^{ \sigma \beta } g^{\gamma \rho} g^{\delta \nu}, \label{nol}
\end{eqnarray}
where, $D_{i}'s\ (i\in\{1,2,3,4,6\}) $ are functions of $\phi$ and $X=\nabla_{\mu}\phi\nabla^{\mu}\phi$. There should be another term in Eq.(\ref{nol}) of the form $D_5 g^{ \mu \eta } g^{  \nu \beta } g^{\gamma \rho} g^{ \delta \sigma}$ \cite{Joshi:2021azw} which is not considered here due to antisymmetry properties of the Riemann tensor.

The form of $ \tilde{H}^{\mu \nu \rho \sigma \alpha \beta \gamma \delta}$ in  Eq.(\ref{nol}) is derived by considering the symmetricity properties of the Riemann tensor and scalar field derivatives ($\nabla_{\mu} A_{\nu}=\nabla_{\nu} A_{\mu}$).
  By introducing a Lagrange multiplier, the Eq.(\ref{S0}) becomes,
\begin{eqnarray}
 S= \int d^{4}x \sqrt{-g} \  \tilde{H}^{\mu \nu \rho \sigma \alpha \beta \gamma \delta} R_{\alpha\beta\gamma\delta} \nabla_{\mu} A_{\nu} \,\nabla_{\rho} A_{\sigma} +\lambda^{\mu}( \nabla_{\mu}\phi-A_{\mu})  \label{S1},
\end{eqnarray}
which further can be written as,
 \begin{eqnarray}
\begin{aligned}
 S=\int d^{4}x \sqrt{-g} \ (L_{1}+L_{2}+L_{3}+L_{4}+L_{6})+\lambda^{\mu} (\nabla_{\mu} \phi-A_{\mu}),\label{simple1}
 \end{aligned}
\end{eqnarray}
where $L_i$s  (i=1,2,3,4,6) are defined as,
\begin{eqnarray}
\begin{aligned}
L_{1}&=D_{1} R  g^{c e} g^{d f}  \nabla_{c} A_{d} \nabla_{e} A_{f},\label{l16} &\\
L_{2}&= D_{2}R g^{c d} g^{e f}  \nabla_{c} A_{d}  \nabla_{e} A_{f},&\\
L_{3}&=D_{3} g^{a c} g^{b e} g^{d f} R_{a b}  \nabla_{c} A_{d} \nabla_{e} A_{f}, \\
L_{4}&=D_{4} g^{a e} g^{c d} g^{b f} R_{a b} \nabla_{c} A_{d}   \nabla_{e} A_{f}, &\\
L_{6}&= D_{6}g^{c a} g^{f b} g^{l e} g^{m d} R_{a b l m}\nabla_{c} A_{d} \nabla_{e} A_{f}.&
\end{aligned}
\end{eqnarray}
\subsubsection{3+1 Decomposition of Lagrangian \texorpdfstring{$L_{1-6}$}{TEXT}}
Now, we decompose Lagrangians $L_{1}$ to $L_{6}$, by using the 3+1 decomposed  relations of  $ \nabla_{c} A_{d}$ in eq.(\ref{R1}), and the curvature tensor provided in Eq.(\ref{sdfe}, (\ref{R6},(\ref{R3}),\ref{R2}) ). Then we separate the higher derivative of the metric and scalar field with the corresponding coefficient, which yields,
\begin{eqnarray} 
\begin{aligned}
L_{1-6} =&  \left(2h^{a b} \mathcal{L}_{n}   K_{a b}+ \mathcal{F}_{1} \right) Z_{*}^2\left(D_{1}+D_{2}\right)+
\left(h^{a b}\mathcal{L}_{n} K_{a b} +\mathcal{F}_{2}\right) Z_{*}^2\left(D_{3}+D_{4}\right)\\&+
h^{a b}\mathcal{L}_{n} K_{a b}Y_{c}Y^{c}
\left(-4D_1-D_3\right)+h^{ab}\mathcal{L}_{n} K_{a b} Z_{*} U\left( -4D_{2}-D_4\right)\\&+\mathcal{L}_{n} K_{a b}Y^{a}Y^{b}\left(-D_{3}-2D_{6}\right)
+\left(2h^{a b} \mathcal{L}_{n} K_{a b}+\mathcal{F}_{1}\right)\left( D_{1} U^{cd} U_{cd}+D_{2}U^2\right) \\&+\mathcal{L}_{n} K_{a b}Z_{*}  U^{ab}\left( -D_4+2D_6\right) 
+ \left(\mathcal{L}_{n} K_{a b}+F_{ab}\right)\left(D_{3} U_{d}^{a} U^{ b d} +D_{4}U U^{a b}\right)\\&+Y^{d}Y_{d}\left(-2D_{1}\mathcal{F}_{1}+D_{3} \mathcal{F}_{2}\right)
+Z_{*}U\left(-2D_{2}\mathcal{F}_{1}+D_{4}\mathcal{F}_{2} \right) \\&+ 2 V_b \left[ Z_{*} Y^{b}\left(D_{3}+D_{4}\right)-D_{3}U_{a}^{b}Y^{a}-D_{4}UY^{b} \right]
+Y^{a}Y^{b}(-F_{ab}D_{3}+2D_{6}\mathcal{F}_{ab}\\&+Z_{*}U^{a b}(-F_{ab}D_{4}-2D_{6}\mathcal{F}_{ab})
-D_{6}\mathfrak{R}_{a b c d}U^{ac}U^{bd}+D_{6}4U^{bc}Y^{a}V_{cab}.\label{ssa}
\end{aligned}
 \end{eqnarray}
In this expression, problematic terms are, 
\begin{itemize}
     \item  $Z_{*}^2 \,$, the second order quadratic derivative of scalar field(QSDS), \item $Z_{*} \,$, linear second order derivative of scalar field (LSDS) \item  ${L}_{n} K_{a b}$, linear second order derivative of metric (LSDM) \item $Z_{*}^2\,{L}_{n} K_{a b}$, mix term of QSDS and LSDM \item $Z_{*}\,{L}_{n} K_{a b}$, mix term of LSDS and LSDM  \end{itemize}
     
 It is easy to see from Eq.(\ref{ssa}) that we can tune the coefficients $D_1-D_6$ in such a way that some problematic terms (QSDS or its mixed term with LSDM, the mixed term of LSDM and LSDS) vanish. Consequently, we obtain the following conditions,
 \begin{eqnarray}
\begin{aligned}
D_1+D_2=&0, \ \ \ \  D_3+D_4=0,\\ -4D_1-D_3=&0,  \ \ \ \ 4D_2+D_4=0, \\ -D_3-2D_6=&0,  \ \ \ \ D_4-2D_6=0.\label{con1}
\end{aligned}
\end{eqnarray}
After solving these equations, we get
\begin{eqnarray}
   D_1=-D_{2}= -\dfrac{D_3}{4}=\dfrac{D_4}{4}=\dfrac{D_{6}}{2}.\label{sigma} 
   \end{eqnarray} 
After applying these conditions, $L_{1-6}$ becomes,
 \begin{eqnarray}
 \begin{aligned}
L_{1-6}=& D_{1}\bigg[\left(U^{cd}U_{cd}-U^{2}\right)\left(2h^{ab}\mathcal{L}_{n}K_{ab}+\mathcal{F}_{1}\right)+4\left(U U^{ab}-U^{a}_{d}U^{db}\right)\left(\mathcal{L}_{n}K_{ab}+F_{ab}\right)\\&+2Z_{*}U\left(\mathcal{F}_{1}+2\mathcal{F}_{2} \right)-4Z_{*}U^{a b}(F_{ab}+\mathcal{F}_{ab})-2
Y^{d}Y_{d}\left(\mathcal{F}_{1}+2 \mathcal{F}_{2}\right)\\&+
 8 V_b \left(U_{a}^{b}Y^{a}-UY^{b}\right)+4
Y^{a}Y^{b}(F_{ab}+\mathcal{F}_{ab})
-2\mathfrak{R}_{a b c d}U^{ac}U^{bd}+8U^{bc}Y^{a}V_{cab} \Bigg]. \label{rip} 
 \end{aligned}
 \end{eqnarray}
 \subsubsection{Ghost Free Lagrangian in Unitary Gauge}
  Note that by virtue of Eq.(\ref{sigma}), the LSDS and LSDM are the only existing problematic terms in the $L_{1-6}$  with the Lagrangian coefficient $D_{1}$. In order to remove LSDS and LSDM, the only option is the  $D_{1}=0$ \cite{Joshi:2019tzr}, a trivial choice.  This theory can also be checked in the unitary gauge to eliminate problematic terms, but there are no other options besides a trivial one.
  
  It is suggested in \cite{Joshi:2021azw} that to remove LSDM, we need to introduce new Lagrangians, which are quadratic in curvature and coupled with the first derivatives of scalar fields. And to remove LSDS, we must introduce coupled Lagrangians of quartic second-order scalar field derivatives. Finally, the following combination of the Lagrangian becomes ghost-free in the unitary gauge,
 \begin{multline}
L_{1-16}=\bigg(D_{1} R  g^{c e} g^{d f} +D_{2}R g^{c d} g^{e f}  
+D_{3} g^{a c} g^{b e} g^{d f} R_{a b} +D_{4} g^{a e} g^{c d} g^{b f} R_{a b}+D_{6}g^{c a} g^{f b} g^{l e} g^{m d} R_{a b l m}\bigg)\nabla_{c} A_{d} \nabla_{e} A_{f}\\+\bigg(D_{7}g^{a b} g^{c d} g^{e f}  R_{a b}R_{c d}+ D_{8}g^{a b} g^{c e} g^{d f}  R_{a b}R_{c d}+ D_{9}g^{a c} g^{b d} g^{e f}  R_{a b}R_{c d}+ D_{10}g^{a c} g^{b e} g^{d f}  R_{a b}R_{c d}\\+ D_{11}g^{c r} g^{q b} g^{d f} g^{s e} g^{a p} R_{a c b d}R_{p q r s}\bigg)A_{e} A_{f}+\bigg(D_{12}g^{ab} g^{cd} g^{ef} g^{pq}+D_{13} g^{ae} g^{dp} g^{cf} g^{qb}\\+D_{14}g^{ac} g^{bd} g^{ep} g^{qf}+D_{15}g^{ab} g^{cd} g^{pf} g^{qe}+D_{16}g^{ab} g^{cq} g^{pf} g^{de} \bigg) \nabla_{a} A_{b}\nabla_{c} A_{d} \nabla_{e} A_{f} \nabla_{p} A_{q}+\lambda^{a} (\nabla_{a} \phi-A_{a}),
 \end{multline}
 Now, we decompose the above Lagrangian $L_{1}$ to $L_{16}$ by using 3+1 formalism and separate the higher derivatives of the scalar field and metric with corresponding coefficient,
\begin{eqnarray}
\begin{aligned}
L_{1-16}=&L_{1-6}+\mathcal{L}_n K_{ab} \mathcal{L}_n K_{cd}\bigg\{ \mathcal{A}^{a} \mathcal{A}^{c} h^{bd}(D_{10} + D_{11})-A_{*}^2 h^{ac} h^{bd}(D_{11}+D_{9})+2 D_{8} \mathcal{A}^{a} \mathcal{A}^{b} h^{cd} \\& -  A_{*}^2 h^{ab} h^{cd}(4 D_{7}+2 D_{8} +D_{9}+D_{10})+ D_{9} \mathcal{A}_{e} \mathcal{A}^{e} h^{bc} h^{ad} + (4 D_{7} + D_{9}) \mathcal{A}_{e} \mathcal{A}^{e} h^{ba} h^{cd}  \bigg\}\\&+\mathcal{L}_{n} K_{a b}\bigg\{ D_{8} \mathcal{F}_{1} \mathcal{A}^{a} \mathcal{A}^{b} +2 A_{*}^2 (- D_{9}  F^{ab} + D_{11}\mathcal{F}^{ab}) - \mathcal{F}_{1} A_{*}^2 h^{ab} (4 D_{7} + D_{8}) - 2 D_{10} \mathcal{A}^{a} A_{*} V^{b} \\& + 2 A_{*}^2 \mathcal{F}_{2} h^{ab} (D_{8}+D_{9}+D_{10})+ 2 D_{9} \mathcal{A}_{c} \mathcal{A}^{c} F^{ab}  + 2 D_{10} \mathcal{A}^{c} \mathcal{A}^{b} F^{a}{}_{c} - 2 D_{11} \mathcal{A}^{c} \mathcal{A}^{b} \mathcal{F}^{a}{}_{c} \\& + 2\mathcal{A}_{c} \mathcal{A}^{c} h^{ab} (2 D_{7} \mathcal{F}_{1}- D_{9} \mathcal{F}_{2})- 2\mathcal{A}^{c} A_{*} h^{ab} V_{c} (D_{10} + 2 D_{8})  + 2 D_{11} \mathcal{A}^{c} A_{*} V_{c}{}^{ab}+ 2 D_{8} \mathcal{A}^{c} \mathcal{A}^{d} F_{cd} h^{ab}\bigg\}\\&+Z_{*}^4\left(D_{12} + D_{13}  + D_{14} + D_{15} + D_{16}\right) + 4 Z_{*}^3 U\left(-4 D_{12}  - 2 D_{15} - D_{16}\right)
\\& +  Z_{*}^2\bigg\{ U^{2}\left(6 D_{12}+ D_{15}\right)- 4 Y_{a} \ Y^{a}\left(-4 D_{13}  + 4  D_{14} - 2 D_{15}- 3 D_{16}\right)
 +2  U_{bc} U^{bc}\left(2 D_{14}+ D_{15}\right)\bigg\}\\&
+Z_{*}\bigg\{-4 D_{12}U^{3}+4 D_{13}  U_{ab} Y^{a} Y^{b}+D_{15}(- 2  U U_{de} U^{de} + U \ Y_{b} Y^{b}) +D_{16}( - U_{b}{}^{d} U^{bc} U_{cd}  \\& + 3 \ U^{c}{}_{c} Y_{b} Y^{b} + 3 U_{bc} Y^{b} Y^{c} )\bigg\}+L_{FDSD}.
\end{aligned}
\end{eqnarray} 
 We have already identified the higher derivatives in $L_{1-6}$ Eq.(\ref{ssa}) and the relation Eq.(\ref{sigma}) among $D_{1}-D_{6}$ to remove some of them. For the new Lagrangian with coefficient $D_{7}-D_{11}$, the problematic terms are quadratic($\mathcal{L}_{n} K_{a b}\mathcal{L}_{n} K_{cd}$) and linear($\mathcal{L}_{n} K_{a b}$) second derivatives of the metric and for $ D_{12}-D_{16}$, these terms are quartic($Z_{*}^4$), cubic($Z_{*}^3$), quadratic($Z_{*}^2$) and linear($Z_{*}$) second order derivatives of scalar field. Where $L_{FDSD}$ contains the first and spatial derivatives of the metric and scalar field. It can be observed that  the higher derivatives $Z_{*}^4$, $Z_{*}^3$ and $Z_{*}^2$ completely vanish for the conditions, 
 \begin{eqnarray}
D_{12} + D_{13}  + D_{14} + D_{15} + D_{16} =0,\nonumber\\
-4 D_{12}  - 2 D_{15} - D_{16}=0,\nonumber\\
2 D_{14}+ D_{15}=0,\\
6 D_{12}  + D_{15}=0,\nonumber\\
-4 D_{13}  - 4  D_{14} - 2 D_{15}- 3 D_{16}=0.\nonumber
\end{eqnarray}  and the solution of these equations are,
\begin{eqnarray}
D_{15}& =-6D_{12}= D_{13}=-2 D_{14}=-\frac{3}{4} D_{16}.\label{de1}
  \end{eqnarray} 
Please note that the aforementioned conditions are obtained solely for simplicity of calculation. They may not the only relations that could give us the desired result and there may exist a more diverse class of relations, of which these may be a special case.

Also, notice that no  sufficient conditions exist for removing $Z_{*}$, $\mathcal{L}_n K_{ab}$ and  $\mathcal{L}_n K_{ab}\ \mathcal{L}_n K_{cd}$. There is no nontrivial relation between the coefficients of Lagrangian  $L_{1-16}$ to remove all the higher derivatives, implying that the theory does not exhibit degeneracy(when the scalar field is the function of spacetime).

If we restrict theory to unitary gauge the nontrivial solution of equations related to higher derivatives  $\mathcal{L}_n K_{ab}$ and  $\mathcal{L}_n K_{ab}\ \mathcal{L}_n K_{cd}$ is possible, they are
\begin{eqnarray}
    4D_{7}+2D_{8}+D_{9}+D_{10}=&0,  \qquad D_{9}+D_{11}=0,
\end{eqnarray}and other equations can be related to other coefficients of Lagrangian
  \begin{eqnarray}
  \begin{aligned}
     2D_{1}=-(4D_{7}+D_{8}),\ \  2D_{1}=D_{9},\ \
    D_{15}=\frac{3D_{1}}{X},\label{de2}  
  \end{aligned}
    \end{eqnarray}
After putting these conditions the finally the Lagrangian takes the form, 
\begin{eqnarray}
    L_{1-16}= D_{1}\bigg[2 A_{*}^{2}  \mathcal{L}_{n}  K^{a b}\left(h_{ab}\mathcal{R}-2\mathcal{R}_{ab}\right)-2\left(h_{ab}\mathcal{R}-2\mathcal{R}_{ab}\right)K^{ab} A_{*}(\mathcal{L}_{n}A_{*})\bigg]+L_{FDSD} \label{fg} 
\end{eqnarray}
the higher derivative finally remove with the two degeneracy conditions (for more details refer to \cite{Joshi:2021azw}), 
   \begin{itemize}
       \item \textbf{Con1}: If the combination $h^{ab}\mathcal{R}-2\mathcal{R}^{ab} $ vanishes.
       \item \textbf{Con2}: If the first condition does not hold, then  $D_{1}=\frac{\mathcal{C}}{X^\frac{3}{2}}.$ 
       \end{itemize}
       This theory can be considered a ghost-free theory in the unitary gauge, analogous to the class of UDHOST theories for these particular choices.
  As a final comment about this theory, the Lagrangian $L_{1-16}$ does not hold any degeneracy condition for the general scalar field, indicating no nontrivial relation exists among the coefficient of Lagrangians to remove higher derivatives. However, in the unitary gauge, a nontrivial solution exists among the coefficient of Lagrangian, indicating the presence of degeneracy.

  In the next section, we shall modify the Lagrangian $L_{1-6}$ by introducing additional coupling of two first-order derivatives of the scalar field.

 \section{Extension of \texorpdfstring{$L_{1-6}$}{TEXT} With Two First Derivative of Scalar Field}  \label{l16fd}
In the previous section, we discussed how one could eliminate ghosts in the Lagrangian $L_{1-6}$ by following \cite{Joshi:2021azw}.  In what follows, we derive another method of getting rid of ghosts by introducing additional coupling $A_{\mu}A_{\nu}$ in Eq.(\ref{S0}). As seen here, this additional coupling does not introduce new higher derivatives but rather generates new Lagrangians. Mathematically, instead of the tensor $ \tilde{H}^{\mu \nu \rho \sigma \alpha \beta \gamma \delta}$ of rank 8, a tensor of rank 10 has to be used, we called it $\tilde{H}_{1} ^{\mu \nu \rho \sigma \alpha \beta \gamma \delta \eta \zeta}$. In this case, we consider the following action,
  \begin{eqnarray}
 S= \int d^{4}x \sqrt{-g} \  \tilde{H}_{1}^{\mu \nu \rho \sigma \alpha \beta \gamma \delta \eta \zeta} R_{\alpha\beta\gamma\delta} \nabla_{\mu} A_{\nu} \,\nabla_{\rho} A_{\sigma}  A_{\eta}\,A_{\zeta}+\lambda^{\mu}( \nabla_{\mu}\phi-A_{\mu})  \label{S3}.
\end{eqnarray} 
where the tensor $\tilde{H}_{1}^{\mu \nu \rho \sigma \alpha \beta \gamma \delta \eta \zeta}$ is given as,
 \begin{eqnarray}
 \begin{aligned}
 \tilde{H}_{1}^{\mu \nu \rho \sigma \alpha \beta \gamma \delta \eta \zeta}= g^{\eta \zeta}\, \tilde{H}^{\mu \nu \rho \sigma \alpha \beta \gamma \delta}+ H_{6} g^{\alpha\gamma} g^{\beta\delta}g^{\mu \nu} g^{\eta \rho}g^{\sigma\zeta}+H_{7} g^{\alpha\gamma} g^{\beta\delta}g^{\mu \rho} g^{\eta \nu}g^{\sigma\zeta}+H_{8} g^{\alpha\gamma} g^{\beta\mu}g^{\sigma\zeta} g^{\nu\delta}g^{\rho\eta}\\+H_{9} g^{\alpha\gamma} g^{\beta\mu}g^{\rho\delta} g^{\nu\eta}g^{\sigma\zeta}+H_{10} g^{\alpha\gamma} g^{\beta\mu}g^{\delta\eta} g^{\nu\zeta}g^{\rho\sigma}+H_{11} g^{\alpha\gamma} g^{\beta\eta}g^{\delta\mu} g^{\rho\nu}g^{\sigma\eta}\\+H_{12} g^{\alpha\gamma} g^{\beta\eta}g^{\delta\zeta} g^{\mu\nu}g^{\rho\sigma}+H_{13} g^{\alpha\gamma} g^{\beta\eta}g^{\mu\sigma} g^{\nu\rho}g^{\delta\zeta}+H_{14} g^{\alpha\mu} g^{\beta\rho}g^{\gamma\eta} g^{\nu\zeta}g^{\rho\sigma}\\H_{15} g^{\alpha\mu} g^{\beta\eta}g^{\gamma\nu} g^{\delta\zeta}g^{\rho\sigma}+H_{16} g^{\alpha\mu} g^{\beta\eta}g^{\gamma\rho} g^{\delta\zeta}g^{\sigma\nu}.\label{exl}
  \end{aligned}
 \end{eqnarray}
 where $H_i$'s are functions of $\phi$ and $X$. To avoid confusion from the previous section, we use the Lagrangian coefficient $H_{i}'s$ instead of $D_{i}'s$, where we have taken $H_5=D_6$ for consistency (since the contribution from $L_5$ was zero owing to the antisymmetric properties of the Riemann tensor). Other terms are obtained by considering all possible couplings, resulting in a total 16 Lagrangians. Eq.(\ref{S1}) can be written in the following form,
 \begin{eqnarray}
\begin{aligned}
 S=\int d^{4}x \sqrt{-g} \ (L^{\prime}_{1}+L_{2}^{\prime}+L_{3}^{\prime}+L_{4}^{\prime}+L_{5}^{\prime}+L_{6}^{\prime}+L_{7}^{\prime}+L_{8}^{\prime}+L_{9}^{\prime}+L_{10}^{\prime}+L_{11}^{\prime}+L_{12}^{\prime}\\+L_{13}^{\prime}+L_{14}^{\prime}+L_{15}^{\prime}+L_{16}^{\prime})+\lambda^{\mu} (\nabla_{\mu} \phi-A_{\mu}),\label{simple}
 \end{aligned}
\end{eqnarray}
where $L_i$s  (i=1,2...6) are defined as,
\begin{eqnarray}
\begin{aligned}
L^{\prime}_{1}&=H_{1}Rg^{ec}g^{ab}g^{df}\,\nabla_{c}A_{d}\nabla_{e}A_{f}\,A_{a}\,A_{b} &\\
L^{\prime}_{2}&= H_{2} R g^{ab}g^{dc}g^{ef}\,\nabla_{c}A_{d}\nabla_{e}A_{f}\,A_{a}\,A_{b},&\\
L^{\prime}_{3}&=H_{3}g^{pc}g^{re}g^{df}g^{ab}R_{pr}\,\nabla_{c}A_{d}\nabla_{e}A_{f}\,A_{a}\,A_{b}, \\
L^{\prime}_{4}&=H_{4}g^{pe}g^{rf}g^{dc}g^{ab}R_{pr}\,\nabla_{c}A_{d}\nabla_{e}A_{f}\,A_{a}\,A_{b},&\\
L_{5}^{\prime}&= H_{5}g^{cp}g^{qf}g^{ds}g^{re}g^{ab} R_{pqrs}\,\nabla_{c}A_{d}\nabla_{e}A_{f}\,A_{a}\,A_{b},&\\
L_{6}^{\prime}&=H_{6}Rg^{cd}g^{ea}g^{fb}\,\nabla_{c}A_{d}\nabla_{e}A_{f}\,A_{a}\,A_{b},&\\
L_{7}&=H_{7}R g^{ce}g^{da}g^{fb}\,\nabla_{c}A_{d}\nabla_{e}A_{f}\,A_{a}\,A_{b},&\\
L_{8}^{\prime}&=H_{8}g^{rp}g^{qc}g^{sd}g^{ea}g^{fb}R_{pqrs}\,\nabla_{c}A_{d}\nabla_{e}A_{f}\,A_{a}\,A_{b},&\\
L_{9}&=H_{9}g^{rp}g^{qc}g^{se}g^{da}g^{fb}R_{pqrs}\,\nabla_{c}A_{d}\nabla_{e}A_{f}\,A_{a}\,A_{b},&\\
L_{10}^{\prime}&=H_{10}g^{rp}g^{qc}g^{sa}g^{db}g^{ef}R_{pqrs}\,\nabla_{c}A_{d}\nabla_{e}A_{f}\,A_{a}\,A_{b},&\\
L_{11}^{\prime}&=H_{11}g^{rp}g^{qa}g^{sc}g^{de}g^{bf}R_{pqrs}\,\nabla_{c}A_{d}\nabla_{e}A_{f}\,A_{a}\,A_{b},&\\
L_{12}^{\prime}&=H_{12}g^{rp}g^{qa}g^{sb}g^{dc}g^{ef}R_{pqrs}\,\nabla_{c}A_{d}\nabla_{e}A_{f}\,A_{a}\,A_{b},&\\
L_{13}^{\prime}&=H_{13}g^{rp}g^{qa}g^{sb}g^{cf}g^{de}R_{pqrs}\,\nabla_{c}A_{d}\nabla_{e}A_{f}\,A_{a}\,A_{b},&\\
L_{14}^{\prime}&=H_{14}g^{cp}g^{qe}g^{ra}g^{db}g^{fs}R_{pqrs}\,\nabla_{c}A_{d}\nabla_{e}A_{f}\,A_{a}\,A_{b},&\\
L_{15}^{\prime}&=H_{15}g^{cp}g^{qa}g^{rd}g^{sb}g^{ef}R_{pqrs}\,\nabla_{c}A_{d}\nabla_{e}A_{f}\,A_{a}\,A_{b},&\\
L_{16}^{\prime}&=H_{16}g^{cp}g^{qa}g^{re}g^{sb}g^{fd}R_{pqrs}\,\nabla_{c}A_{d}\nabla_{e}A_{f}\,A_{a}\,A_{b},
\end{aligned}
\end{eqnarray}
\subsection*{3+1 Decomposition of Lagrangian \texorpdfstring{$L^{\prime}_{1-16}$}{TEXT}}
 Now, we decompose the above Lagrangian $L_{1}^{\prime}$ to $L_{16}^{\prime}$ by 3+1 formalism,
 \begin{eqnarray}
L_{1-16}^{\prime}=L^{\prime}_{QSDS}+L^{\prime}_{LSDS}+L^{\prime}_{LSDM}+L^{\prime}_{FDSD},\label{rhed}
 \end{eqnarray}
 where $L^{\prime}_{QSHS}$ contains QSDS and its mixed term with LSDM, $L^{\prime}_{LSHS}$ contains LSDS and its mixed term with LSDM, $L^{\prime}_{LSDM}$ contains LSDM, and $L^{\prime}_{FDSD},$ contains the first derivative of the metric and scalar field, and spatial derivative terms. The full 3+1 decomposition of the Lagrangians ($L_{1}^{\prime}$ to $L_{16}^{\prime}$) is given in Appendix B.
 Here, we write the higher derivative step by step. First, the $L^{\prime}_{QSHS}$,
\begin{eqnarray}
\begin{aligned}
L^{\prime}_{QSDS}=
Z_{*}^2\Biggl[A_{*}^2\biggr\{\,(H_{3} + H_{4} + H_{8} + H_{9}+H_{10} + H_{11} + H_{12} + H_{13} ) \mathcal{F}_{2}-(H_1 +  H_2 + H_6 + H_7 ) \mathcal{F}_{1} \\- h^{ab} \mathcal{L}_n K_{ab}\,(2 H_1+2 H_{2}+H_{3}+H_{4}+2 H_{6}+2 H_{7}+H_{8}+H_{9} +H_{10}+ H_{11}+H_{12}+H_{13}) \biggr\}\\ + \mathcal{A}^{a} \mathcal{A}^{b}\biggl\{ (H_{12} + H_{13}) F_{ab} + (H_{15}+H_{16}) \mathcal{F}_{ab}+( H_{12}+H_{13}+H_{15}+H_{16}) \mathcal{L}_n K_{ab}\biggr\}\\+ A_{*} \mathcal{A}^{a} V_{a}(H_{10}+H_{11}+2 H_{12}+2 H_{13}) + \mathcal{A}_{a}\, \mathcal{A}^{a} \
\biggl\{(H_{1}+H_{2}) \mathcal{F}_{1}\\ - ( H_{3}+H_{4}) \mathcal{F}_{2}+ (2 H_{1} + 2 H_{2} + H_{3} + H_{4}) \
(h^{ab} \mathcal{L}_n K_{ab})\biggr\}\Biggr]\label{HHA}
\end{aligned}
\end{eqnarray}
It can be observed from the above to Eq.(\ref{HHA}) to eliminate QSDS, the following condition on the coefficient of Lagrangian is obtained,
 \begin{eqnarray}
 \begin{aligned}
 H_{3} + H_{4} + H_{8} + H_{9}+H_{10} + H_{11} + H_{12} + H_{13}&= 0,\\ H_{1} + H_{2} + H_{6} + H_{7} &= 0,\\
 2 H_1+2 H_{2}+H_{3}+H_{4}+2 H_{6}+2 H_{7}+H_{8}+H_{9} +H_{10}+ H_{11}+H_{12}+H_{13}&= 0,\\
 H_{12} + H_{13}&= 0,\\
H_{15} + H_{16}&= 0,\\
 H_{12} + H_{13}+H_{15} + H_{16} &= 0,\\
H_{10} + H_{11} + 2 H_{12} + 2 H_{13} &= 0,\\
H_1 + H_{2} &= 0,\\
H_3 + H_4 &= 0,\\
2 H_{1} + 2 H_2 + H_3 + H_4 &= 0.\label{conQSHS}
 \end{aligned}
 \end{eqnarray}
 The solution to these equations is,
\begin{equation}
    H_{2}=-H_{1},\quad H_{3}=-H_{4}, \quad H_{7}=-H_{6},\quad H_{9}=-H_{8},\quad H_{11}=-H_{10},\quad H_{13}=-H_{12},\quad H_{16}=-H_{15}\label{conds3}
\end{equation}
 The existence of a solution indicates the presence of a degeneracy for removing QSDS for the set of simultaneous linear equations above. Using these relations, we can remove QSDS and its mixed term with LSDM for the general case (for any metric and space-time function of the scalar field). From now, the independent coefficients in the full Lagrangian are $\{ H_1,  H_4, H_6, H_8, H_{10}, H_{12}, H_{14},H_{15}\}$.
Next, we plan to remove LSDS and its mixed term with LSDM. 
 \begin{multline}
L_{LSDS}^{\prime}=Z_{*}\Bigg[A_{*}^2 \Bigl\{(H_4 + H_8) F^{ab} U_{ab} + (H_{14} + H_{15} + 2 H_5) \mathcal{F}^{ab} \
U_{ab} + (-2 H_1+H_{6}) \mathcal{F}_{1} U -  (H_{10}+ 2 H_{12} +  H_{4}) \mathcal{F}_{2} U \\+ (-H_{14} + H_{15} + H_{4}-2 H_{5}+ H_{8}) U^{ab} \mathcal{L}_n K_{ab}- (4 H_1-H_{10}-2 H_{12}-H_{4}- 2 H_6) h^{ab} U \mathcal{L}_n K_{ab}\
\Biggr\}\\ + \mathcal{A}^{a}A_{*} \Biggl\{(H_{10} - 2 H_8)V^{b} U_{ab} -  (H_{14} + 2 H_{15}) V_{a}{}^{bc} U_{bc}+ (H_{10} + 4 H_{12}) V_{a} U + 2 H_{14} \mathcal{F}_{ab} Y^{b}\Biggl\}\\ + \mathcal{A}^{a} \mathcal{A}_{a}\Biggl\{ (2 H_1 \mathcal{F}_{1} + H_4  \mathcal{F}_{2}) U-  (H_4 F^{bc}+ 2 H_5 \mathcal{F}^{bc})U_{bc}  - (H_4- 2 H_5) U^{bc} \mathcal{L}_n K_{bc}+ (4 H_1-  H_4) Uh^{bc} \mathcal{L}_n K_{bc} \Biggl\}\\+ \mathcal{A}^{a}\mathcal{A}^{b} 
\Bigl\{(U_{ab} \bigl(-H_{6} \mathcal{F}_{1} + H_8 \mathcal{F}_{2}) - (H_{10}F_{a}{}^{c}+ H_{14}\mathcal{F}_{a}{}^{c})  U_{bc} - (2 H_{12} F_{ab} + \mathcal{F}_{ab} H_{15})  U\\ -  H_{15} \mathfrak{R}_{a c b d}U^{cd}- ( H_{10}- H_{14} ) U_{a}{}^{c}\mathcal{L}_n K_{bc} -(2 H_{12}+H_{15}) U \mathcal{L}_n K_{ab}-(2H_{6}+H_{8}) U_{ab} h^{cd}\mathcal{L}_n K_{cd} \Bigr\}\Bigg]\label{lkj}
\end{multline}
Similar to the previous case, we obtain the following linear equations to remove LSDS,
   \begin{eqnarray}
\begin{aligned}
 H_{4} + H_{8} = 0,&\\
  H_{14} + H_{15} +2 H_{5} = 0,&\\
  -2H_{1}+H_{6}=0&\\
  H_{10} + 2 H_{12} + H_{4} = 0,&\\
  H_{10} - 2 H_{8} = 0,&\\
  H_{14} + 2 H_{15} = 0,&\\
  H_{10} + 4 H_{12} = 0,
\label{conds2}
\end{aligned}
\end{eqnarray}
and to remove mixed terms of LSDM and LSDS, 
\begin{eqnarray}
\begin{aligned}
 -H_{14} + H_{15} + H_{4} - 2 H_{5} + H_{8} = 0,&\\
  4 H_{1} - H_{10} - 2 H_{12} - H_{4} - 2 H_{6} = 0,&\\
  H_{4} - 2 H_{5} = 0,&\\
   4 H_{1} - H_{4} = 0,&\\
    H_{10} - H_{14} = 0,&\\
  2H_{12} + H_{15} = 0,&\\
  2 H_{6} +H_{8} = 0.
 \label{conds1}
\end{aligned}
\end{eqnarray}  
 No nontrivial solution exists for the Eq's (\ref{conds2}-\ref{conds1}), i.e., no relation exists among the $ H_{i}'s$ to switch off LSDS. Here, the only possible solution is the trivial one, that all the $ H_{i}'s=0$. Since all equations are simultaneous, the number of independent equations is more than the number of variables. 

To conclude, for any general space-time and scalar field, we cannot remove the LSDS and its mixed term with LSDM, implying that no degeneracy condition exists. Therefore, we are not looking at the possibility of eliminating LSDM. 
In the next section, we will check whether any degeneracy condition exists in the unitary gauge.

\section{Analysis in Unitary Gauge}\label{unitary}
Following our analysis of the previous section, we note that for any case, the number of independent equations can be reduced, then the  Eq's (\ref{conds2}-\ref{conds1}) have nontrivial solutions. For this, the unitary gauge is the best choice. The following condition characterizes a unitary gauge,
 \begin{equation}
     \phi(x,t)=\phi_{o}(t).
 \end{equation} 
 For more details about unitary gauge, refer to \cite{Gleyzes:2014dya, Gleyzes:2014qga, Langlois:2015cwa}. Since the scalar field only has time-dependence in this gauge, it is clear that $\mathcal{A}^{a} $ vanishes here,  and the other variables that appeared in eq.(\ref{R1}) now become,
 \begin{eqnarray}
   U_{a b}= - A_{*} K_{a b},\ \ \ Y_{b}=0,\ \ Z_{*}=\mathcal{L}_{n}A_{*}.\label{lkoi}
 \end{eqnarray} Using these relations, Eq.(\ref{lkj})becomes in unitary gauge,
 \begin{multline}
L_{LSDS}^{\prime}=-\mathcal{L}_{n}A_{*}\,A_{*}^3\Bigg[(H_4 + H_8) F^{ab} K_{ab} + (H_{14} + H_{15} + 2 H_5) \mathcal{F}^{ab} \
K_{ab} + (-2 H_1+H_{6}) \mathcal{F}_{1} K -  (H_{10}+ 2 H_{12} +  H_{4}) \mathcal{F}_{2} K \\
+ (-H_{14} + H_{15} + H_{4}-2 H_{5}+ H_{8}) K^{ab} \mathcal{L}_n K_{ab}- (4 H_1-H_{10}-2 H_{12}-H_{4}- 2 H_6) h^{ab} K \mathcal{L}_n K_{ab}\Bigg]
\end{multline}
To remove $\mathcal{L}_{n}A_{*}$, we need following relations,
 \begin{eqnarray}
\begin{aligned}
4 H_{1} - H_{10} - 2 H_{12} - H_{4} - 2 H_{6} = 0,&\\
  -H_{14} + H_{15} + H_{4} - 2 H_{5} + H_{8} = 0,&\\
  H_{4} + H_{8} = 0,&\\
  H_{14} + H_{15} +2 H_{5} = 0,&\\
  -2H_{1} +H_{6} = 0,&\\
  H_{10} + 2 H_{12}+H_{4} = 0.&\\
\end{aligned}
\end{eqnarray}
The solution to these equations is,
\begin{eqnarray}
\begin{aligned}
 H_4 = -H_8,\quad H_6 = 2H_{1},\quad H_{10} = -2H_{12}+H_{8},\quad H_{14} = -2{H}_{5},\quad H_{15} = 0.\label{sasaki}
\end{aligned}
\end{eqnarray}
After imposing Eq (\ref{sasaki}, $L^{\prime}_{LSDM}$ and $L^{\prime}_{FDSD}$ remain in the total Lagrangian $L^{\prime}_{1-16}$ in Eq.(\ref{rhed}. In unitary gauge, we obtain the following form:
 \begin{multline}
L^{\prime}_{LSDM}+L^{\prime}_{FDSD}=A_{*}^4 \Bigl[H_{1}\mathcal{F}_{1}(K^{2}- K_{ab} K^{ab})- H_8 F^{bc}( K_{b}{}^{d} K_{cd}+K K_{bc}) + H_{12}\mathcal{F}_2( K^2- K_{ab} K^{ab})\\ + H_5 \mathfrak{R}_{a b c d} K^{bc} K^{de} +\ \mathcal{L}_n K_{cd} \Bigl\{ (-2H_{1}+H_{12}) \
h^{cd} K_{be} K^{be}+(2H_{1}-  H_{12}) h^{cd} K^{2} -  H_8( K_{b}{}^{d} K^{bc} - K K^{cd}) \Bigr\}\Bigr]\label{lklk}
\end{multline}
 In this expression, one can see that from the three independent variables $\{H_{1}, H_{5}, H_{8}, H_{12}\}$, only the LSDM is present with $H_{1}, $ $H_{8}$ and $H_{12}$ (second line). If we impose  $H_{8}=0$ and $H_{12}=2H_{1}$, then the final form of action becomes independent of any higher derivative term,
\begin{eqnarray}
L^{\prime}_{GF}=A_{*}^4 \Bigl[H_{5}\mathfrak{R}_{a b c d} K^{bc} K^{de}+H_{1}(K^{2}- K_{ab} K^{ab})(\mathcal{F}_{1}+2\mathcal{F}_{2})\Bigr].\label{youj}
\end{eqnarray}
Where $L^{\prime}_{GF}$ stands for ghost-free Lagrangian. The calculations above indicate that the full Lagrangian $L^{\prime}_{1-16}$ can be made free from any instability in the unitary gauge.
The above-derived condition for removing higher derivative terms are not a necessary condition, because the first solution set Eq.(\ref{conds3}) for $\phi(x,t)$, and other solution set Eq.(\ref{sasaki}) and, $H_{8}=0$ and $H_{12}=2H_{1}$ is derived in unitary gauge.

In  $L^{\prime}_{GF}$, the independent variables are $H_{1}$, and $ H_{5}$. 
First, if we take $H_{1}=0$,  in the initial Lagrangian system $L_{1-16}^{\prime}$, all coefficients in the Lagrangian are set to zero, except  $H_{5}$ and $H_{14}$, which are related by the relation $ H_{14} = -2 H_{5}$ (note, once again, that $H_{5}$ and $H_{14}$ are a general functions of $\phi$ and $X$). Then, the final form of action is ghost free in unitary gauge.
Second, If we taken $H_{1 }\neq 0$, then in the initial Lagrangian system, by virtue of Eq.(\ref{conds3}) and Eq.(\ref{sasaki}) the final ghost free Lagrangian contains $L^{\prime}_{1}$,  $L^{\prime}_{2}$ $L^{\prime}_{6}$ $L^{\prime}_{7}$ $L^{\prime}_{10}$ $L^{\prime}_{11}$,  $L^{\prime}_{12}$ $L^{\prime}_{13}$.

One might notice that the theory in \textsection{III} is ghost free in unitary gauge for specific cases (see Con1 and Con2 in \textsection{III C}) using relations listed in Eqs.(\ref{de1},\ref{de2}) among Lagrangian coefficients. However, our case does not require additional conditions to be a ghost-free theory in the unitary gauge; only the relation between the Lagrangian coefficients is needed, indicating that the degeneracy of the Lagrangian is more general. Additionally, in  \textsection{III}, the ghost-free Lagrangian in unitary gauge contains linear and quadratic curvature terms out of the combination of 16 different Lagrangians. Still, our case contains only two Lagrangians, which are linear in curvature.

Consider a  interesting case, if $H_{5}=\frac{1}{X_{U}^{2}}$, where, $X_{U}$ is value of $X$ in unitary gauge, then $H_{5}=\frac{1}{A_{*}^{4}}$. In this case, Eq.(\ref{youj}) takes the following form, 
\begin{eqnarray}
    L^{\prime}_{GF}=\mathfrak{R}_{a b c d} K^{bc} K^{de}
\end{eqnarray}
This equation contains only the metric derivatives that can yield two degrees of freedom, i.e., GR-like theory.

\section{ Checking for FRW Background}\label{frwback}
We will now describe the above calculation for the flat, closed, and open FLRW universes as an example, whose metric can be expressed as,
 \begin{equation}
     ds^2=-N(t)\,dt^2+a(t)^{2}\left(\frac{dr^2}{1-\mathcal{K}r^2} +r^{2}d\theta^2+r^2sin^2\theta d\phi^{2}\right).\label{metri},
 \end{equation}
written in spherical polar coordinates $r$, $\theta$, $\phi$, where  $\mathcal{K} $ is the spatial curvature, $N(t)$ is the lapse function, and $a(t)$ is the scale factor.
For FLRW metric the Eq.(\ref{youj}) takes the following form,
\begin{eqnarray}
L^{\prime FRW}_{GF}=6\left(6H_{1}(\phi,X_{U})+H_{5}(\phi,X_{U})\right) \,{\frac {\left(\dot{a}^{2}+\mathcal{K} N^2 \right)\dot{a}^{2} \dot{\phi}^{4}}{ a^{4} \ N^{8}}}\label{frw}
\end{eqnarray}
where, $X_{U}=-\frac{\dot{\phi}^{2}}{N^2}$. We are taking $6H_{1}(\phi,X_{U})+H_{5}(\phi,X_{U})=\mathcal{H}(\phi,X_{U})$.
        The $L^{\prime FRW}_{GF}$ has the same mathematical form as obtained in \cite{Joshi:2021azw} if we choose $6\mathcal{H}(\phi,X_{U})=\frac{D^{`}(\phi,X_{U})}{X_{U}}$ and $\mathcal{K}=0$, where $D^{'}=-36D_{7}-12D_{9}$. 
    The structure of both theories is different for the general scalar field $(\phi(x,t))$ as well as the unitary gauge $ (\phi(t))$. But in the case of FRW background, both attend similar final forms for a particular value of $H_{5}$.
Here, our action is,
\begin{eqnarray}
  S= \int d^{4}x \sqrt{-g} \left[\frac{R}{2\kappa}-\frac{1}{2}g^{\mu\nu}\nabla_{\mu}\phi\nabla_{\nu}\phi-V(\phi)+L^{\prime FRW}_{GF}\right],\
  \label{general}
\end{eqnarray}
where $\kappa^{2}=\frac{1}{M_{p}^{2}}$, $M_{p}$ is the four-dimensional Planck mass, and $V (\phi)$ is the scalar potential.  
For the metric eq.(\ref{frw}), the action (in unitary gauge) becomes,
\begin{equation}
  S_{FRW}= \int d^{4}x a^{3}N \left[\frac{\ddot{a}-\dot{a}^2}{a^{2}N^{2}}+\frac{\dot{\phi}^{2}}{2N^2}-V(\phi)+6\mathcal{H}\Big(\phi,X_{U}\Big)\,{\frac {\left(\dot{a}^{2}+\mathcal{K} N^2 \right)\dot{a}^{2} \dot{\phi}^{4}}{ a^{4} \ N^{8}} }\right].\end{equation}
From this action, we obtain the following  equations of motion,
   \begin{eqnarray}
\begin{aligned}
-\left(H^{4}+\frac{\,H\mathcal{K}}{9\,a^{2}}\right) \mathcal{H}(\phi,X_{U})\,\dot{\phi}^{4}+ \left(H^{3}+\frac{\,H\mathcal{K}}{2\,a^{2}}\right)\bigg(16\frac{\partial \mathcal{H}(\phi,X_{U})}{\partial X_{U}}
\dot{\phi}^{5}\ddot{\phi}-8\frac{\partial \mathcal{H}(\phi,X_{U})}{\partial \phi}  \dot{\phi}^{5}\\-32 \,\mathcal{H}(\phi,X_{U})
\dot{\phi}^{3}\ddot{\phi} \bigg) -24\left( H^{2}+\frac{\mathcal{K} }{6 \,a^{2}}\right) \mathcal{H}(\phi,X_{U})\dot{H}\dot{\phi}^{4}\\+\left(\frac{1}{2}\dot{\phi}^{2}-\,V \left( \phi \right)\right)+
 3\frac{H^{2}}{\kappa}+2\frac{\dot{H}}{\kappa}+\frac{\mathcal{K}}{a^{2}\kappa}=0,
\end{aligned}
\end{eqnarray}
\begin{eqnarray}
\begin{aligned}
-6\,\left( 7\,H^{4}+\frac{5\,H^{2}\mathcal{K}}{a^{2}}\right)\mathcal{H}(\phi,X_{U})
\dot{\phi}^{4} +12\,\left( H^{4}+\frac{H^{2}\mathcal{K}}{a^{2}}\right)
\frac{\partial \mathcal{H}(\phi,X_{U})}{\partial X_{U}}\dot{\phi}^{6}\\-(V \left( \phi \right) +\frac{1}{2}\dot{\phi}^{2})+3\, \frac{H^{2}}{\kappa}+\frac{3\mathcal{K}}{a^{2}\kappa}=0,
\end{aligned}
\end{eqnarray}
\begin{eqnarray}
\begin{aligned}
 \left(H^{5}+\frac{\,H^{3}\mathcal{K}}{3\,a^{2}}\right) \left(36\frac{\partial H_5(\phi,X_{U}}{\partial X_{U}} \dot{\phi}^{5}-72 H_{5}(\phi,X_{U})\dot{\phi}^{3}\right)+\left(H^{4}+\frac{H^{2}\mathcal{K}}{a^{2}}\right)\bigg( 12 \frac{\partial^2 H_{5}(\phi,X_{U})}{\partial \phi\, \partial X_{U}}\dot{\phi}^6\\
 -24 \frac{\partial^2 H_{5}(\phi,X_{U})}{\partial X_{U}^{2}}\dot{\phi}^6\ddot{\phi}+108\frac{\partial H_{5}(\phi,X_{U})}{\partial X_{U}}\dot{\phi}^{4}\ddot{\phi}-18\frac{\partial H_{5}(\phi,X_{U})}{\partial \phi}\dot{\phi}^{2}\dot{\phi}^{4}-72H_{5}(\phi,X_{U})\ddot{\phi}\bigg)\\
 +\left(H^{3}+\frac{ H\mathcal{K}}{2\,a^{2}}\right)\bigg(48\,\frac{\partial H_{5}(\phi,X_{U})}{\partial X_{U}} \dot{\phi}^{5}  \dot{H}-96H_{5}(\phi,X_{U})\dot{\phi}^{3}\dot{H}\bigg)\\+3\, \dot{\phi} H - \frac{\partial V   \left( \phi\right)}{\partial \phi} -\,\ddot{\phi} =0.
\end{aligned}
\end{eqnarray}
Putting $\mathcal{K}=0$, one can obtain the FLRW equations for the flat case.

\section{de Sitter Inflation Conditions}\label{dsinflation}
In this section, we intend to find the set of conditions on the higher derivative interaction terms that may retain the inflationary de Sitter phase as seen in the standard model of cosmology. Please note that we are not working with any particular inflationary potential in mind. However, we assume that the potential allows for a slow-roll inflationary period. Considering the two terms we arrived at,
\begin{align}\label{corr}
L_1'+L_2'+L_{5}^{\prime}+L_6'+L_7'+L_{10}'+L_{11}'+L_{12}+L_{13}+L_{14}^{\prime}&= H_{1}Rg^{ec}g^{ab}g^{df}\,\nabla_{c}A_{d}\nabla_{e}A_{f}\,A_{a}\,A_{b}\nonumber\\&+H_{2} R g^{ab}g^{dc}g^{ef}\,\nabla_{c}A_{d}\nabla_{e}A_{f}\,A_{a}\,A_{b}\nonumber\\&+H_{5}g^{cp}g^{qf}g^{ds}g^{re}g^{ab}R_{pqrs}\,\nabla_{c}A_{d}\nabla_{e}A_{f}\,A_{a}\,A_{b}\nonumber\\&+H_{6}Rg^{cd}g^{ea}g^{fb}\,\nabla_{c}A_{d}\nabla_{e}A_{f}\,A_{a}\,A_{b}\nonumber\\&+H_{7}R g^{ce}g^{da}g^{fb}\,\nabla_{c}A_{d}\nabla_{e}A_{f}\,A_{a}\,A_{b}\nonumber\\&+H_{10}g^{rp}g^{qc}g^{sa}g^{db}g^{ef}R_{pqrs}\,\nabla_{c}A_{d}\nabla_{e}A_{f}\,A_{a}\,A_{b}\nonumber\\&+H_{11}g^{rp}g^{qa}g^{sc}g^{de}g^{bf}R_{pqrs}\,\nabla_{c}A_{d}\nabla_{e}A_{f}\,A_{a}\,A_{b}\nonumber\\&+H_{12}g^{rp}g^{qa}g^{sb}g^{dc}g^{ef}R_{pqrs}\,\nabla_{c}A_{d}\nabla_{e}A_{f}\,A_{a}\,A_{b}\nonumber\\&+H_{13}g^{rp}g^{qa}g^{sb}g^{cf}g^{de}R_{pqrs}\,\nabla_{c}A_{d}\nabla_{e}A_{f}\,A_{a}\,A_{b}\nonumber\\&+H_{14}g^{cp}g^{qe}g^{ra}g^{db}g^{fs}R_{pqrs}\,\nabla_{c}A_{d}\nabla_{e}A_{f}\,A_{a}\,A_{b}
\end{align}
along with the degeneracy condition in the unitary gauge: $
 H_2 = -H_1,\ H_7 = -H_{6},\ H_{11} = -H_{10},\ H_{13} = -H_{12},\ H_6 = 2H_{1},\ H_{10} = -4H_{1},\ H_{14} = -2{H}_{5}, \ \textnormal{and}\  H_{12}=2H_{1}$. Through dimensional analysis, we can arrive at a general analytic expression for $H_p$, where $p\in\{1,2,5,6,7,10,11,12,13,14\}$ as:
\begin{equation}\label{const}
    H_p=\frac{\zeta_p\phi^m X^n}{M_P^{m+4n+8}}
\end{equation}
where $\zeta_p$ is a dimensionless coupling constant corresponding to the coefficient $H_p$ and $m$, $n$ are integers. The coefficient $H_p$ could also be a linear combination of any number of such coefficients. Imposing a de Sitter inflation phase, we can constrain $\zeta_p$ by demanding that the additional terms are perturbative additions to the Einstein-Hilbert term. A slow-rolling de Sitter inflation phase also imposes additional constraints on the theory, like constant Hubble parameter $\Dot{H}\to0$, small $\Dot{\phi}$ and $\Ddot{\phi}\to0$, that may be used to reduce the number of terms added to the equations of motion owing to (\ref{corr}). Applying these, we find that the correction to the Einstein equations of motion is:
\begin{align}\label{eomterm}
    H_1(\phi,X)[2R_{pb}R_{qc}\partial_a\phi\partial^a\phi\partial^b\phi\partial^c\phi+g_{pq}R_{bc}R\partial_a\phi\partial^a\phi\partial^b\phi\partial^c\phi-2R_{p\ b}^{\ d\ e}R_{qdce}\partial_a\phi\partial^a\phi\partial^b\phi\partial^c\phi\nonumber\\
    -2g_{pq}R_{a\ b}^{\ e\ f}R_{cedf}\partial^a\phi\partial^b\phi\partial^c\phi\partial^d\phi-2R_{ab}R_{pcqd}\partial^a\phi\partial^b\phi\partial^c\phi\partial^d\phi-R_{qb}R\partial_a\phi\partial^a\phi\partial^b\phi\partial_p\phi\nonumber\\
    +2R_{bdce}R_{q\ a}^{\ d\ e}\partial^a\phi\partial^b\phi\partial^c\phi\partial_p\phi-R_{pb}R\partial_a\phi\partial^a\phi\partial^n\phi\partial_q\phi+2R_{bdce}R_{p\ a}^{\ d\ e}\partial^a\phi\partial^b\phi\partial^c\phi\partial_q\phi\nonumber\\
    -2R_a^cR_{bc}\partial^a\phi\partial^b\phi\partial_p\phi\partial_q\phi+R_{ab}R\partial^a\phi\partial^b\phi\partial_p\phi\partial_q\phi]+H_5(\phi,X) R_{bdce}R_{p\ q}^{\ d\ e}\partial_a\phi\partial^a\phi\partial^b\phi\partial^c\phi
\end{align}
where we have replaced $\nabla_a\to\partial_a$ due to the derivative acting on the scalar field. Since we have worked in the unitary gauge ($\phi\equiv\phi(t)$) so far, it makes sense to impose the same here. We can use the following relations applicable to FLRW spacetime backgrounds:
\begin{equation}
    R=12\left(\frac{\Dot{a}}{a}\right)^2,\quad R_{00}=-3\left(\frac{\Dot{a}}{a}\right)^2,\quad R^i_{\ 0j0}=-\left(\frac{\Dot{a}}{a}\right)^2\delta^i_{\ j},\quad R^0_{\ i0j}=\left(\frac{\Dot{a}}{a}\right)^2\delta^i_{\ j}.
\end{equation}
Using these relations, the terms in (\ref{eomterm}) become
\begin{align}
    3(40H_1+H_5)H^4\Dot{\phi}^4;\quad &\text{for timelike indices $p$ and $q$, and}\nonumber\\
    -48H_1H^4\Dot{\phi}^4;\quad &\text{for spacelike indices $p$ and $q$.}
\end{align}
First, we note that $H_5$ contributes only to the timelike part of the equations of motion, i.e. it affects only the energy density. Also, setting $H_1=0$ renders the spacelike part 0, implying that in such a case there are no pressure corrections to the energy-momentum tensor. Nevertheless, both terms are proportional to $H^4\Dot{\phi}^4$. Explicitly imposing the condition that the scale of these terms must be infinitesimal compared to the Einstein tensor in order to preserve a stable de Sitter inflation phase, we obtain the following relation:
\begin{align}
    \zeta &\ll \frac{M_P^{m+4n+10}}{\phi^m\Dot{\phi}^{2n+4}H^2}\nonumber\\
    \implies \zeta &\ll \left(\frac{1}{\epsilon}\right)^{n+2}\left(\frac{M_P}{\phi}\right)^m \left(\frac{M_P}{H}\right)^{2n+6}\label{constr3}
\end{align}
where $\zeta$ is used to represent some linear combination of $\zeta_1$ and $\zeta_5$. In the last step, we have introduced the slow-roll parameter $\epsilon$ to eliminate $\Dot{\phi}$ from the final expression. From various inflationary models, the limits on $H$ vary from $\approx 10^{-24}$ GeV (below which the universe is too cool for big bang nucleosynthesis) and $\approx 10^{-14}$ (below which EW baryogenesis is difficult to achieve) \cite{Jiang2016} to the upper limit $\approx 10^{14}$ GeV (set by the data \cite{Ade_2016}) compared to the Planck scale ($\approx 10^{19}$ GeV).

From the inequality above, it is straightforward to see that in order for the condition to be satisfied for all $\phi\ll M_P$, the coupling constant $\zeta$ could assume a wide range of relatively small, but relevant, values. For $\phi\gg M_P$, the constraint on $\zeta$ becomes increasingly tighter as $\phi$ becomes larger to the point where because of the coefficient, the term could become irrelevant in other epochs ($\phi$ represents the classical background of the scalar field which assumes different values in different cosmological epochs). 

\section{Unitarity Analysis}\label{unitarity}
We shall, now, try to find whether for conditions similar to the standard cosmological inflation, the theory remains unitary up to the Planck scale. Equivalently, we can use the condition of unitarity to impose constraints on the coupling parameter (similar to the analysis for $R^2\Phi^2$ coupling in \cite{Panda:2022esd}) found in the coefficient of the higher derivative coupling term introduced in theory.

To this end, we first expand the given fields and quantize their respective perturbations. Since we've established broadly the conditions for which a stable inflationary phase occurs in (\ref{constr3}), we can expand the metric around a de Sitter background as follows:
\begin{equation}\label{hquant}
    g^{ab}=\Bar{g}^{ab}+\kappa h^{ab},
\end{equation}
where $\Bar{g}^{ab}$ represents the de Sitter background metric and $h^{ab}$ represents perturbations to it which shall be quantized for scattering analysis. Similarly, we expand the scalar field as
\begin{equation}
    \phi(x,t)=\varphi(x,t)+\phi'(x,t).
\end{equation}
Here, both the background field $\varphi$ and the perturbations $\phi'$ to be quantized are taken initially to be functions of both space and time. Unitary gauge conditions would later be imposed on the background field to obtain accurate results. Quantization of fields around curved backgrounds has been laid out beautifully in \cite{Birrell:1982ix, Parker:2009uva}. Our ability to use the Bunch-Davies vacuum \cite{Birrell:1982ix, Parker:2009uva} allows us the privilege of quantizing fields in a straightforward manner. Further, we can use techniques like the local momentum space method \cite{Birrell:1982ix,Parker:2009uva,panda2023local}to perform a Fourier transform and to move the calculations to the momentum space.

Since defining a local momentum space involves working along a geodesic where the covariant derivatives reduce to partial derivatives along with some curvature corrections, we can obtain approximate physical momenta-based information, though it is restricted only to the geodesic. However, considering the homogeneity of spacetime, we could assume that the same results could very well describe physics in neighboring regions as well. Also, the curvature corrections about the geodesic can be expressed in terms of the connection, Hubble parameter ($H$), and its derivatives. Given that we're working with a slow-roll inflationary background, and that interested in scales approaching the Planck scale, it is feasible for us to ignore these corrections.

Expanding (\ref{corr}) up to the third order in perturbations while restricting to the first order in $\kappa$ and keeping the slow-roll conditions in mind ($\Dot{\varphi}$ is small; $\Ddot{\varphi}\to0$), we find that the higher-derivative terms only provide corrections to the $\phi'^2h_{ab}$-vertex. Keeping in mind that we're working in the ultraviolet limit, the correction to the aforementioned 3-vertex is given as: 
\begin{align}\label{vertexfn}
    V_{\phi'^2h_{ab}}=&i\kappa [\{-(2H_1+H_5)(\partial\varphi\cdot p_1)(\partial\varphi\cdot p_2)(p_1\cdot p_2)+(2H_1-H_5)(\partial{\varphi}\cdot p_1)^2(p_1\cdot p_2)\nonumber\\&+2H_1(\partial\varphi)^2(p_1\cdot p_2)\}(p_{1_d}p_{2_e}+p_{1_e}p_{2_d})
    +2\{(2H_1+H_5)(\partial\varphi\cdot p_1)^2(p_1\cdot p_2)+(2H_1+H_5)(\partial\varphi)^2(p_1\cdot p_2)^2\nonumber\\&
    +(H_5-2H_1)(\partial\varphi\cdot p_1)(\partial\varphi\cdot p_2)(p_1\cdot p_2)\}(p_{1_d}p_{1_e})\nonumber\\&
    +\{(6H_1+H_5)p_1+(2H_1-H_5)p_2\}\cdot\partial\varphi(p_1\cdot p_2)^2(\partial\varphi_d p_{1_e}+\partial\varphi_e p_{1_d})\nonumber\\&
    -4H_1(p_1\cdot p_2)^3\partial\varphi_d\partial\varphi_e
    -H_1\{2(\partial\varphi)^2(p_1\cdot p_2)+4(\partial\varphi\cdot p_1)^2\}\eta_{de}(p_1\cdot p_2)^2]
\end{align}
where the momenta correspond only to the scalar field perturbations and are taken to be directed towards the vertex. Note that the coefficients $H_{\{1,5\}}$ don't contribute to the perturbative expansion, given the slow-roll conditions that we choose to work with. Therefore, in (\ref{vertexfn}), $H_{\{1,5\}}\equiv H_{\{1,5\}}(\varphi,\Bar{X})$, where $\Bar{X}=\partial_a\varphi\partial^a\varphi$. Now, owing to the same arguments as before, we can use the Minkowski background propagators for both metric and scalar field perturbations, which are as follows:
\begin{align}\label{prop}
    \left<h^{ab}h_{cd}\right>&=i\frac{(\delta^a_c\delta^b_d+\delta^b_c\delta^a_d)-\eta^{ab}\eta_{cd}}{2k^2} + \mathcal{O}(\lambda),\\
    \left<\phi'\phi'\right>&=\frac{i}{k^2},
\end{align}
where $\lambda$ is the gauge parameter introduced to the graviton propagator by the gauge fixing term, and doesn't contribute on-shell, such as for scattering amplitude calculations that we're about to perform. For a $\phi'^2\to\phi'^2$ process, the terms in (\ref{corr}) contribute only to the processes involving $h_{ab}$-exchange in all three channels: $s$, $t$, and $u$. 

Since the scattering amplitude is independent of gauge choice, we can simply work in the center of mass frame to proceed with the calculations, while keeping the factors of the form $\partial_a\varphi$ and $\partial_a\varphi\partial^a\phi'$ as is. For such a case, with the center of mass energy being $E$, the Mandelstam variables are $s=2p_1\cdot p_2=2p_3\cdot p_4=4E^2$, $t=-2p_1\cdot p_3=-2p_2\cdot p_4=-2E^2(1-\cos \theta)$, and $u=-2p_1\cdot p_4=-2p_2\cdot p_3=-2E^2(1+ \cos \theta)$. Once we obtain the amplitude in the center of mass frame, we can finally impose slow-roll conditions for which $\partial_a\varphi\to\Dot{\varphi}$ and $\partial_a\varphi\partial^a\phi'\to\Dot{\varphi}\partial_0\phi'$ or $\partial_a\varphi k^a\to\Dot{\varphi}k^0$ in the Fourier space.  We also assume that, up to a constant factor, $k^0\approx E$ as well, and find that the correction to the scattering amplitude is:
\begin{align}\label{finalamp}
    i\mathcal{M}_{HD}=&i\left(\kappa^2\Dot{\varphi}^4 E^{10}\right)[\left(1152 H_1^2-180 H_1 H_5-138 H_5^2\right)+\left(-2088 H_1^2-744 H_1 H_5-180 H_5^2\right) \cos \theta \nonumber\\
    & +\left(416 H_1^2+160 H_1 H_5+260 H_5^2\right) \cos^2 \theta+\left(48 H_1^2-16 H_1 H_5+4 H_5^2\right) \cos^3 \theta \nonumber\\
    & +\left(-32 H_1^2-44 H_1 H_5+6 H_5^2\right) \cos^4 \theta+\left(-8 H_1^2-8 H_1 H_5-2 H_5^2\right) \cos^5 \theta]
\end{align}
One simplifying case could be where we assume that $H_1=H_5=D$ where the amplitude correction becomes:
\begin{equation}\label{finalampred}
    i\mathcal{M}_{HD}=i\left(\kappa^2\Dot{\varphi}^4 E^{10}\right)D^2[834-2962 \cos \theta+836 \cos ^2 \theta+36 \cos ^3 \theta -70 \cos ^4 \theta-18 \cos ^5 \theta]
\end{equation}

From (\ref{finalamp}) and (\ref{finalampred}), we see how the tree-level unitarity of the theory up to the Planck scale is highly dependent on the form of the coefficients $H_{\{1,5\}}$. Also peculiar, is the fact that these corrections don't introduce a correction to the amplitude that diverges in the collinear ($\theta\to\{0,\pi\}$) limit.

\subsection{An Example}\label{example}
It is straightforward to see that if in (\ref{const}) we set $m=0$ and $n=-1$, the amplitude in (\ref{finalampred}) becomes,
\begin{equation}
    i\mathcal{M}_{HD}\propto \zeta^2\left(\frac{E^{10}}{M_P^{10}}\right)
\end{equation}
where we have substituted $\kappa=2/M_P$ and assumed that $\zeta_{\{1,5\}}=\zeta$, in light of $H_1=H_5=D$. We see that the amplitude is safe from UV unitarity violations up to the scale $\sim\frac{M_P}{\zeta^{1/5}}$, up to residual violations. Besides the constraint on the coupling paramters introduced on the basis of the requirement for a de Sitter expansion phase (\ref{constr3}): $\zeta\leq\frac{M_P^4}{H^4}$; we can now impose a constraint based on unitarity: $\zeta\leq1$. This ensures that the tree-level unitarity violations happen near or above the Planck scale. Note that these parameter constraints are highly dependent on the form of $D$ and, therefore, the aforementioned constraint should not be considered universal.

\subsection{Renormalizability and Unitarity}
We can also choose a form of $D$ based on power-counting arguments to ensure renormalizability of the theory (up to the order of perturbation under consideration) and, then, impose unitarity constraints on the coefficients, which would help provide us with a well-behaved theory. Power-counting arguments, based on the mass dimensions of the coefficients of interaction terms, can be used to obtain a crude estimate for the renormalizability of the theory, though more comprehensive analyses (similar to that performed in \cite{PhysRevD.16.953,Abe:2017abx,Abe:2018rwb,Abe:2020ikj,Larin:2019zhb}) may be needed to ensure renormalizability. According to this method, a theory is said to be super-renormalizable for constant coefficients of positive mass dimensions, renormalizable for constant coefficients of zero mass dimensions, and non-renormalizable for constant coefficients of negative mass dimensions. For reference, the example in section \ref{example} would render the theory non-renormalizable.

For power-counting arguments to be applicable, the fields must be quantized and of mass dimension 1. As such, any conditions can only be put forward after we have quantized the perturbation $h_{\mu\nu}$ in (\ref{hquant}). Then, the condition for renormalizability based on (\ref{const}) could be expressed as: $m+4n+9=0$, while that for super-renormalizability is: $m+4n+9<0$. Considering a simple example for renormalizability, assume $m=-5$ and $n=-1$, such that the amplitude becomes:
\begin{equation}
    i\mathcal{M}_{HD}\propto \zeta^2\left(\frac{E^{10}}{\varphi^{10}}\right)
\end{equation}
i.e. whether the UV unitarity violations happen is now up to the epoch within the theory, under consideration; specifically the magnitude of the scalar field background. The violation scale in this case is $\sim\frac{\varphi}{\zeta^{1/5}}$. For a scalar background that ranges from small, say for reheating era, to an intermediate value below the Planck scale, we'd need $\zeta\ll1$ to avoid unitarity violations up to the Planck scale. During inflation, when the scalar background could assume values in the trans-Planckian regime, as per the previous section, the unitarity constraints would dictate that the coupling parameter be $\zeta\gg1$, which doesn't agree with the condition for de Sitter expansion, implying that there is no overlap between unitarity and renormalizability here. Similar analysis for different combinations of $m$ and $n$ along the straight line $m+4n+9=0$ can be performed.

Similarly for a super-renormalizable theory, we assume $m=-6$ and $n=-1$ such that:
\begin{equation}
    i\mathcal{M}_{HD}\propto \zeta^2 M_P^2\left(\frac{E^{10}}{\varphi^{12}}\right)
\end{equation}
where the constant coefficient now has a mass dimension of 2. Here, the violation scale is $\sim\frac{\varphi^{6/5}}{\zeta^{1/5}M_P^{1/5}}$, and the condition to avoid violations is $\varphi\geq\zeta^{1/6}M_P$. If the scalar background assumes any value below the Planck scale, we'd need $\zeta\ll1$. This time, however, approaching a small scalar background (reheating regime) results in $\zeta\to0$, meaning that in order to preserve unitarity, we'd need that the higher derivative coupling terms be ignored throughout, thus eliminating small scalar background as an acceptable range for these terms to contribute to the theory. As for a trans-Planckian background, $\zeta$ could assume any value as long as the condition $\zeta\leq\left(\frac{\varphi}{M_P}\right)^6$ holds. 

So far, we have explicitly avoided including $\Dot{\varphi}$ in the final amplitude through conscious choices of $n$. There may also be cases where $\Dot{\varphi}$ is retained in the final expression. Owing to slow-roll conditions, positive powers of $\Dot{\varphi}$ can be helpful in ensuring no UV unitarity violations are caused by the higher-derivative terms introduced in this paper, while negative powers could doom the theory far below the Planck scale.

In fact, for $n\leq-2$, we note that $\Dot{\varphi}$ appears in negative powers in the amplitude. To put a reasonable constraint, we can substitute for the slow-roll parameter $\epsilon$, which in turn introduces negative powers of $H$ into the expression. $\zeta$ would effectively have to be zero in order to offset such a contribution to the scattering amplitude. So, to ensure that the theory remains unitary, we impose the constraint that $n\geq-1$ (applicable to both renormalizable and super-renormalizable theories). 


\section{Conclusion}\label{concl}
This work systematically studies existing degeneracy conditions for an action containing curvature (in linear order) coupled with the first- and second-order derivatives of the scalar field. Considering the properties of Riemann tensors and scalar field derivatives, we determine the 16 possible Lagrangians for such couplings. Using the 3+1 decomposition technique, we separate the higher derivatives (QSDS, LSDS, LSDM, and coupling between them) in the 16 different Lagrangians. Linear algebraic relations  can be established between the coefficients of these terms, which can be used as constraints to help eliminate the higher derivative terms from the equation of motion. The existence of the solution to these equations indicates the presence of degeneracy.

There is no solution for the general space-time metric except the trivial one (the Lagrangian coefficients are all zero), which implies that the higher derivatives are always present.
In unitary gauge, a nontrivial solution to the system of simultaneous linear algebraic equations exists.
We obtain the ghost-free combination of Lagrangian in unitary gauge as written explicitly in (\ref{corr}), subjected to the constraints on coefficients added below the equation. 

Testing our formalism for the FRW universe, having imposed all the degeneracy conditions, we obtain second-order equations of motion. This model's background cosmology will be examined as a future project. For now, we only look at the conditions for which a slow-rolling de Sitter inflation phase is expected. The condition on the coupling parameter is mentioned in (\ref{constr3}), and using that and imposing that the coupling is supposed to be small, we find that the scalar field background for this theory could assume values up to the trans-Planckian range, provided that the coefficient of the higher-derivative interaction term doesn't become infinitesimal, rendering the term itself irrelevant to the other epochs.

Further, we inspect the tree-level scattering amplitude for the theory for possible unitarity violations below the Planck scale. This helps us narrow down the form of the coefficient function $D$ and also helps constrain the coupling parameter $\zeta$. We also vaguely introduced renormalizability constraints based on crude power-counting techniques and found conditions, for specific cases, for which an overlap between unitarity and perturbative renormalizability is expected. These conditions, however, are sensitive to the nature of the theory and need to be evaluated on a case-by-case basis.

So far, we have only included up to second-order derivatives of the scalar field in this theory. Degeneracy conditions involving third- and higher-order derivatives can also be verified using the analysis presented above, which we shall leave as future work. 

\section*{ACKNOWLEDGEMENT}
This work is partially funded by DST (Govt. of India), Grant No. SERB/PHY/2021057. Calculations are performed using xAct packages \cite{martin-garcia,Brizuela:2008ra} of Mathematica.
 
\section*{Supplementary Material}
\subsection{Possible Combination}
The structure of  $\tilde{H}^{\mu \nu \rho \sigma \alpha \beta \gamma \delta}$ in terms of the metric depends on the symmetries of the post factor $R_{\alpha\beta\gamma\delta} \nabla_{\mu} A_{\nu} \,\nabla_{\rho} A_{\sigma}  A_{\eta}\, A_{\zeta} $\\(see eq.\ref{S3}). Many possible combinations can exist for this tensor. Since the metric indices are symmetric, the antisymmetric property of the Riemann tensor and the symmetric property of the second-order derivatives of the scalar field makes many of them zero or similar. Some examples include,
\begin{enumerate}
    \item First, the combination comes with the 
$g^{\alpha\beta}...........$
is always vanishes. The same happens with $g^{\gamma\delta}...........$
\item If the first two or last two indices of the Riemann tensor get contracted with the indices of the two second derivatives of scalar field, it always gives a zero result.
\item For $\tilde{H}_{1}^{\mu \nu \rho \sigma \alpha \beta \gamma \delta \eta \zeta}$, if first two or last two indices of the Riemann tensor get contracted with indices of two first-order derivatives of scalar field, the result vanishes.
\end{enumerate}
Other than these, possible terms such that $g^{\alpha\gamma}g^{\beta\delta}$ (i.e. terms involving a Ricci Scalar)
\begin{eqnarray}
\begin{aligned}
&g^{\alpha  \gamma}g^{\beta  \delta} g^{\mu  \nu} g^{\rho  \sigma} g^{\eta  \zeta} \\ &g^{\alpha  \gamma}g^{\beta  \delta} g^{\mu  \nu} g^{\rho  \zeta}g^{\sigma  \eta},\\&g^{\alpha  \gamma}g^{\beta  \delta} g^{\mu  \rho}g^{\nu  \sigma}g^{\eta  \zeta},\\&g^{\alpha  \gamma}g^{\beta  \delta} g^{\mu  \rho}g^{\nu  \eta}g^{\sigma  \zeta},\label{SKR}
  \end{aligned}
\end{eqnarray}
One can find all the terms of this category by exchanging $\eta\longleftrightarrow \zeta$ and  $ \mu\longleftrightarrow \{\nu,\rho,\sigma\}$, but only one pair of indices at a time.
Among all possible combinations, only four in eq.(\ref{SKR}) give different forms; others are similar to these four.
Similar to the term with common factor $g^{\alpha\gamma}$ (i.e. terms involving a Ricci Tensor), the terms with different 3+1 decomposed forms are given as,
\begin{eqnarray}
\begin{aligned}
 &g^{\alpha\gamma}g^{ \beta \rho} g^{ \mu \nu } g^{\delta \sigma} g^{\eta\zeta}\\&  g^{\alpha\gamma}g^{\mu \beta} g^{\delta \rho} g^{\nu \sigma}g^{\eta\zeta}\\& g^{\alpha\gamma}g^{\beta\mu}g^{\sigma\zeta} g^{\nu\delta}g^{\rho\eta}\\& g^{\alpha\gamma} g^{\beta\mu}g^{\rho\delta} g^{\nu\eta}g^{\sigma\zeta}\\& g^{\alpha\gamma} g^{\beta\mu}g^{\delta\eta} g^{\nu\zeta}g^{\rho\sigma}\\& g^{\alpha\gamma} g^{\beta\eta}g^{\delta\mu} g^{\rho\nu}g^{\sigma\eta}\\& g^{\alpha\gamma} g^{\beta\eta}g^{\delta\zeta} g^{\mu\nu}g^{\rho\sigma}\\& g^{\alpha\gamma} g^{\beta\eta}g^{\mu\sigma} g^{\nu\rho}g^{\delta\zeta}
\end{aligned}
\end{eqnarray}
Among all possible combinations, only 8 different forms exist after decomposition in 3+1 formalism; others are similar to those eight.
The terms with all contracted indices (i.e. terms involving a Riemann tensor)
\begin{eqnarray}
\begin{aligned}
&g^{  \mu \alpha} g^{ \sigma \beta } g^{\gamma \rho} g^{\delta \nu}g^{\eta\zeta}
 &\\& g^{\alpha\mu} g^{\beta\rho}g^{\gamma\eta} g^{\nu\zeta}g^{\rho\sigma}\\& g^{\alpha\mu} g^{\beta\eta}g^{\gamma\nu} g^{\delta\zeta}g^{\rho\sigma}\\& g^{\alpha\mu} g^{\beta\eta}g^{\gamma\rho} g^{\delta\zeta}g^{\sigma\nu}
\end{aligned}
\end{eqnarray} For this case, four independent forms of Lagrangian are possible.
\subsection{3+1 decomposed Lagrangian} 
Here, the full expression of the 3+1 decomposition of all 16 Lagrangians is given. These expressions are written in such a way that, in the first bracket, we write the QSDS term and its coupling with LSDM; in the second bracket, we write the LSDQ term and its coupling with LSDM; and in the third bracket, we write the LSDM term alone.
\begin{multline}
  L^{\prime}_{1}=  H_{1}\Bigg[Z_{*}^2 \bigl(F_{1}\ \mathcal{A}_{c} \mathcal{A}^{c}- F_{1} \mathcal{A}_{*}^2 +  - 2 \mathcal{A}_{*}^2 h^{ab}  \mathcal{L}_n K_{ab} + 2 \mathcal{A}_{c} \mathcal{A}^{c}\ h^{ab}  \mathcal{L}_n \ K_{ab}\bigr) \\+ h^{ab} \mathcal{L}_n K_{ab}\bigl(2 \ \mathcal{A}_{e} \mathcal{A}^{e}\ U_{cd} U^{cd}-2 \mathcal{A}_{*}^2 \ U_{cd} U^{cd}  + 4 \ A_{*}^2 \ Y_{c} Y^{c} - 4 \ \mathcal{A}_{e} \mathcal{A}^{e}\ Y_{c}Y^{c}\bigr)\\+F_{1} \bigl(\mathcal{A}_{e} \mathcal{A}^{e}\ U_{cd} U^{cd} -A_{*}^2 \ U_{cd} U^{cd} + 2 A_{*}^2 \ Y_{c} Y^{c} - 2 \mathcal{A}_{e} \mathcal{A}^{e}\ Y_{c} Y^{c}\bigr) \Bigg]
\end{multline}
\begin{multline}
   L^{\prime}_{2}=  H_{2}\Bigg[ Z_{*}^2 \bigl(- F_{1} A_{*}^2 + F_{1} \ \mathcal{A}_{c} \mathcal{A}^{c} - 2 A_{*}^2 h^{ab} \mathcal{L}_nK_{ab} + 2 \ \mathcal{A}_{c} \mathcal{A}^{c}  h^{ab}\mathcal{L}_nK_{ab}\bigr)\\ + Z_{*} U \bigl(2 F_{1} A_{*}^2 - 2 F_{1} \ \mathcal{A}_{c}\mathcal{A}^{c}+ 4 A_{*}^2 h^{ab}\mathcal{L}_nK_{ab} - 4 \mathcal{A}_{c} \mathcal{A}^{c} \ h^{ab}\mathcal{L}_nK_{ab}\bigr)\\  +h^{ab}\mathcal{L}_n K_{ab}\bigl(-2 A_{*}^2 U^2 + 2 \mathcal{A}_{c} \mathcal{A}^{c}\ U^2\bigr) - F_{1} A_{*}^2 U^2 + F_{1}\ \mathcal{A}_{c} \mathcal{A}^{c}\ U^2\Bigg]
\end{multline}
\begin{multline}
   L^{\prime}_{3}=  H_{3}\bigg[ Z_{*}^2 \bigl( A_{*}^2 \mathcal{F}_{2}-\mathcal{A}_{b} \mathcal{A}^{b} \mathcal{F}_{2} -   A_{*}^2 h^{bd} \mathcal{L}_nK_{bd} +  \mathcal{A}_{b} \mathcal{A}^{b} h^{dp} \mathcal{L}_n K_{dp}\bigr)+Z_{*} \bigl( 2  \mathcal{A}_{b} \mathcal{A}^{b} V^{d} Y_{d}-2  A_{*}^2 V^{b} Y_{b}\bigr)\\+   \mathcal{L}_n K_{cd}\bigl(A_{*}^2 Y^{c} Y^{d}+ A_{*}^2 h^{dc} Y_{b} Y^{b}-  A_{*}^2 U_{b}{}^{c} U^{bd}-\mathcal{A}_{b} \mathcal{A}^{b} Y^{d} Y^{c}+\mathcal{A}_{b} \mathcal{A}^{b} U_{d}{}^{d} U^{dc} - \mathcal{A}_{b} \mathcal{A}^{b} h^{cd} Y_{d} Y^{d}\bigr)\\-  A_{*}^2 F^{bd} U_{b}{}^{p} U_{dp} +  \mathcal{A}_{b} \mathcal{A}^{b} F^{dp} U_{d}{}^{r} U_{pr} -   A_{*}^2 \mathcal{F}_{2} Y_{b} Y^{b} +  A_{*}^2 F^{bd} Y_{b} Y_{d} + 2  A_{*}^2 V^{b} U_{bd} Y^{d} \\+  \mathcal{A}_{b} \mathcal{A}^{b} \mathcal{F}_{2} Y_{d} Y^{d} \ -   \mathcal{A}_{b} \mathcal{A}^{b} F^{dp} Y_{d} Y_{p} - 2  \mathcal{A}_{b} \mathcal{A}^{b} V^{d} U_{dp} Y^{p} \bigg]
\end{multline}
\begin{multline}
   L^{\prime}_{4}=  H_{4}\bigg[  Z_{*}^2 \bigl( A_{*}^2 \mathcal{F}_{2}-\mathcal{A}_{b} \mathcal{A}^{b} \mathcal{F}_{2} -   A_{*}^2 h^{bc} \mathcal{L}_n K_{bc} +  \mathcal{A}_{b} \mathcal{A}^{b} h^{cp} \mathcal{L}_n K_{cp}\bigr) \
+ Z_{*} \bigg\{ A_{*}^2 F^{bc} U_{bc} -A_{*}^2 \mathcal{F}_{2} U\\ -   \mathcal{A}_{b} \
\mathcal{A}^{b} F^{cp} U_{cp} +  \mathcal{A}_{b} \mathcal{A}^{b} \mathcal{F}_{2} U - 2  A_{*}^2 V^{b} Y_{b} + 2  \mathcal{A}_{b} \mathcal{A}^{b} V^{c} Y_{c} +  \mathcal{L}_nK_{cp}\big( A_{*}^2 U^{pc} +  A_{*}^2 h^{cp} U \\ -   \mathcal{A}_{b} \mathcal{A}^{b} U^{cp} - \mathcal{A}_{b} \mathcal{A}^{b} h^{cp} U \big)\bigg\}+ \mathcal{L}_n K_{cp} U U^{pc}\big( \mathcal{A}_{b} \mathcal{A}^{b} -A_{*}^2)-  A_{*}^2 F^{bc} U_{bc} U \\+  \mathcal{A}_{b} \mathcal{A}^{b} F^{cp} U_{cp} \
U + 2  A_{*}^2 V^{b} U Y_{b} - 2  \mathcal{A}_{b} \mathcal{A}^{b} \
V^{c} U Y_{c} \bigg]
\end{multline}
\begin{multline}
  L^{\prime}_{5}=  H_{5}\bigg[ Z_{*} \bigl(2  A_{*}^2 \mathcal{F}^{bp} U_{bp}-2\mathcal{A}_{b} \mathcal{A}^{b} \mathcal{F}^{pq} U_{pq} - 2 A_{*}^2 U^{bp} \mathcal{L}_n K_{bp} + 2  \mathcal{A}_{b} \mathcal{A}^{b} U^{pq} \
\mathcal{L}_n K_{pq}\bigr) \\ +\mathcal{L}_n K_{pq}\bigl( 2  A_{*}^2 Y^{q} Y^{p} - 2  \mathcal{A}_{b} \mathcal{A}^{b} Y^{p} Y^{q}\bigr)+ A_{*}^2\ \mathfrak{R}_{a b c d}U^{bp} U^{qr} -   \mathcal{A}_{b} \mathcal{A}^{b}\ \mathfrak{R}_{a b c d} U^{pq} \
U^{rs}\\ + 4  A_{*}^2 V_{b}{}^{pq} U_{pq} Y^{b} + 2  A_{*}^2 \mathcal{F}^{bp} \
Y_{b} Y_{p} - 4  \mathcal{A}_{b} \mathcal{A}^{b} V_{p}{}^{qr} U_{qr} Y^{p} - 2  \
\mathcal{A}_{b} \mathcal{A}^{b} \mathcal{F}^{pq} Y_{p} Y_{q}\bigg]
\end{multline}
\begin{multline}
  L^{\prime}_{6}=  H_{6}\bigg[  Z_{*}^2 \big(-  \mathcal{F}_{1} A_{*}^2 - 2  A_{*}^2 h^{ab} \mathcal{L}_n K_{ab}\big) + Z_{*} \big(  \mathcal{F}_{1} A_{*}^2 U- \mathcal{F}_{1} \mathcal{A}^{a} \mathcal{A}^{b} U_{ab} + 2  \mathcal{F}_{1} \mathcal{A}^{a} A_{*} \ Y_{a} + 2  A_{*}^2 U h^{ab} \mathcal{L}_n K_{ab} \\+ 4 \mathcal{A}^{a} A_{*}Y_{a} h^{bd}  \mathcal{L}_n K_{bd} - 2  \mathcal{A}^{a} \mathcal{A}^{b}U_{ab} h^{de} \
 \mathcal{L}_n K_{de}\big)+\mathcal{L}_n K_{de}(2  \mathcal{A}^{a} \mathcal{A}^{b} h^{de} U_{ab} U - 4  \mathcal{A}^{a} A_{*} h^{ed} U Y_{a})\\+ \mathcal{F}_{1} \mathcal{A}^{a} \mathcal{A}^{b} U_{ab} U - 2  \mathcal{F}_{1} \mathcal{A}^{a} A_{*} U \
Y_{a} \bigg]
\end{multline}
\begin{multline}
     L^{\prime}_{7}=  H_{7}\bigg[Z_{*}^2 \bigl(-  \mathcal{F}_{1} A_{*}^2 - 2  A_{*}^2 h^{ab} \mathcal{L}_n K_{ab}\bigr) + Z_{*} \bigl(2 \mathcal{F}_{1} \mathcal{A}^{a} A_{*} Y_{a} + 4  \mathcal{A}^{a} A_{*} h^{bc} Y_{a} \mathcal{L}_nK_{bc}\bigr)\\+ h^{cd}\mathcal{L}_n K_{cd}\bigl(2 A_{*}^2 Y_{a} Y^{a} + 2  \mathcal{A}^{a} \mathcal{A}^{b}  U_{a}{}^{e} U_{be} - 2  \mathcal{A}^{a} \mathcal{A}^{b}  Y_{a} Y_{b} - 4  \mathcal{A}^{a} A_{*} U_{ab} Y^{b}\bigr)\\+ \mathcal{F}_{1} \mathcal{A}^{a} \mathcal{A}^{b} U_{a}{}^{c} U_{bc} +  \mathcal{F}_{1} A_{*}^2 Y_{a} Y^{a} - \mathcal{F}_{1} \mathcal{A}^{a} \mathcal{A}^{b} Y_{a} Y_{b} - 2  \mathcal{F}_{1} \mathcal{A}^{a} A_{*} U_{ab} Y^{b}\bigg]
\end{multline}
\begin{multline}
   L^{\prime}_{8}=  H_{8}\bigg[  Z_{*}^2 \bigl( A_{*}^2 \mathcal{F}_{2} -A_{*}^2 h^{cd} \mathcal{L}_n K_{cd}\bigr) + Z_{*} \bigg\{ A_{*}^2 F^{cd} U_{cd} +  \mathcal{A}^{c}\mathcal{A}^{d} \mathcal{F}_{2} U_{cd} - 2  \mathcal{A}^{c} A_{*} \mathcal{F}_{2} Y_{c} - 2  A_{*}^2 V^{c} Y_{c}\\ +\mathcal{L}_n K_{ef}\big( A_{*}^2 U^{ef}  + 2  \mathcal{A}^{c} A_{*} h^{fe} Y_{c} -\mathcal{A}^{c} \mathcal{A}^{d} h^{ef} U_{cd}\big)\bigg\}+ \mathcal{L}_n K_{ef}U^{ef}\big( \mathcal{A}^{c} \mathcal{A}^{d} U_{cd}- 2  \
\mathcal{A}^{c} A_{*} Y_{c}\big)\\+ \mathcal{A}^{c} \mathcal{A}^{d} F^{ef} U_{cd} U_{ef}- 2  \mathcal{A}^{c} A_{*} F^{de} U_{de} \
Y_{c} + 4  \mathcal{A}^{c} A_{*} V^{d} Y_{c} Y_{d} - 2  \mathcal{A}^{c} \mathcal{A}^{d} V^{e} \
U_{cd} Y_{e}\bigg]
\end{multline}
\begin{multline}
     L^{\prime}_{9}=  H_{9}\bigg[ Z_{*}^2 \bigr( A_{*}^2 \mathcal{F}_{2} -   A_{*}^2 h^{cd} \mathcal{L}_n K_{cd}\bigl)+ 2Z_{*} \bigr(\mathcal{A}^{c} A_{*} V^{d} U_{cd}-\mathcal{A}^{c} A_{*} \mathcal{F}_{2} Y_{c}- A_{*}^2 V^{c} Y_{c} +  \mathcal{A}^{c} A_{*} h^{de} Y_{c} \mathcal{L}_n K_{de}\bigl)\\+ \mathcal{L}_nK_{ef}\bigl( A_{*}^2 Y^{e} Y^{f} - 2  \mathcal{A}^{c} A_{*} U_{c}{}^{e} Y^{f} + \mathcal{A}^{c} \mathcal{A}^{d} U_{c}{}^{e} U_{d}{}^{f}-\mathcal{A}^{c} \mathcal{A}^{d} h^{ef} Y_{c} Y_{d} \bigr)+ \mathcal{A}^{c} \mathcal{A}^{d} F^{ef} U_{ce} U_{df} \\- 2  \mathcal{A}^{c} \mathcal{A}^{d} V^{e} U_{de} Y_{c}-\mathcal{A}^{c} A_{*} F^{de} U_{ce} Y_{d}+A_{*}^2 F^{cd} Y_{c} Y_{d}+  \mathcal{A}^{c} \mathcal{A}^{d} \mathcal{F}_{2} Y_{c} Y_{d} + 2  \mathcal{A}^{c} A_{*} V^{d}Y_{c} Y_{d}- \mathcal{A}^{c} A_{*} F^{de} U_{cd} Y_{e}\bigg]
\end{multline}
\begin{multline}
    L^{\prime}_{10}=  H_{10}\bigg[ Z_{*}^2 \bigl( A_{*}^2 \mathcal{F}_{2} -   \mathcal{A}^{c} A_{*} V_{c} -   A_{*}^2 h^{cd} \
\mathcal{L}_n K_{cd}\bigr) + Z_{*} \bigg\{ \mathcal{A}^{c} A_{*} V^{d} U_{cd} -   A_{*}^2 \mathcal{F}_{2}U -   \mathcal{A}^{c} \mathcal{A}^{d} F^{f}{}_{c} U_{df} \\+  \mathcal{A}^{c} A_{*} V_{c} U -   \mathcal{A}^{c} A_{*} \mathcal{F}_{2} Y_{c}- A_{*}^2 V^{c} Y_{c} + \mathcal{A}^{c} A_{*} F^{d}{}_{c} Y_{d} +  \mathcal{A}^{c} \mathcal{A}^{d} V_{c} Y_{d}+  \mathcal{L}_n K_{df}\big(A_{*}^2 h^{df} U\\+  \mathcal{A}^{f} A_{*} Y^{d} -   \mathcal{A}^{c} \mathcal{A}^{d} U_{c}{}^{f} +  \mathcal{A}^{c} A_{*} h^{df} Y_{c}\big)\bigg\}+\mathcal{L}_n K_{df}\big( \mathcal{A}^{c} \mathcal{A}^{d} U_{c}{}^{f} U-\mathcal{A}^{f} A_{*} U Y^{d} -\mathcal{A}^{c} A_{*} h^{df} U Y_{c}\big)\\-\mathcal{A}^{c} A_{*} V^{d} U_{cd} U +  \mathcal{A}^{c} \mathcal{A}^{d} F^{f}{}_{c} U_{df} U+  \mathcal{A}^{c} A_{*} \mathcal{F}_{2} U Y_{c} + A_{*}^2 V^{c} U Y_{c} -\mathcal{A}^{c} A_{*} F^{d}{}_{c} U Y_{d}-\mathcal{A}^{c} \mathcal{A}^{d} V_{c} U Y_{d}\bigg]
\end{multline}
\begin{multline}
     L^{\prime}_{11}=  H_{11}\bigg[Z_{*}^2 \bigl( A_{*}^2 \mathcal{F}_{2} -   \mathcal{A}^{c} A_{*} V_{c} -   A_{*}^2 h^{cd} \mathcal{L}_n K_{cd}\bigr) + Z_{*} \big(\mathcal{A}^{c} A_{*} F_{c}{}^{d} Y_{d}-\mathcal{A}^{c} A_{*} \mathcal{F}_{2} Y_{c} -   A_{*}^2 V^{c} Y_{c}\\ +  \mathcal{A}^{c} \mathcal{A}^{d} V_{c} Y_{d} +  \mathcal{A}^{c} A_{*} Y^{d} \mathcal{L}_n K_{cd}\big)+ \mathcal{L}_n K_{ef}\big( A_{*}^2 h^{fe} Y_{c} Y^{c}-\mathcal{A}^{f} A_{*} U_{d}{}^{e} Y^{d}- \mathcal{A}^{c} \mathcal{A}^{f} Y_{c} Y^{e}\\ +\mathcal{A}^{c} \mathcal{A}^{e} U_{c}{}^{d} U_{d}{}^{f} -\mathcal{A}^{c} A_{*} h^{ef} U_{cd} Y^{d}\big)-\mathcal{A}^{c} A_{*} V^{d} U_{c}{}^{e} U_{de}+\mathcal{A}^{c} \mathcal{A}^{d} F_{c}{}^{e} U_{d}{}^{f} U_{ef}\\-A_{*}^2 \mathcal{F}_{2} Y_{c} Y^{c} +  \mathcal{A}^{c} A_{*} V^{d} Y_{c} Y_{d} +  \mathcal{A}^{c} A_{*} \mathcal{F}_{2} U_{cd} Y^{d} +  A_{*}^2 V^{c} U_{cd} Y^{d}+  \mathcal{A}^{c} A_{*} V_{c} Y_{d} Y^{d}\\ -   \mathcal{A}^{c} \mathcal{A}^{d}F_{c}{}^{e} Y_{d} Y_{e} -   \mathcal{A}^{c} A_{*} F_{c}{}^{d} U_{de} Y^{e} -\mathcal{A}^{c} \mathcal{A}^{d} V_{c} U_{de} Y^{e}\bigg]
\end{multline}
\begin{multline}
    L^{\prime}_{12}=  H_{12}\bigg[ Z_{*}^2 \bigl( \mathcal{A}^{c} \mathcal{A}^{f} F_{cf} +  A_{*}^2 \mathcal{F}_{2} - 2  \mathcal{A}^{c} A_{*} V_{c} +\mathcal{A}^{c} \mathcal{A}^{f} \mathcal{L}_n K_{cf} -   A_{*}^2 h^{cf} \mathcal{L}_n K_{cf}\bigr)\\ + Z_{*} \bigl(4  \mathcal{A}^{c}A_{*} V_{c} U-2  A_{*}^2 \mathcal{F}_{2} U-2  \mathcal{A}^{c} \mathcal{A}^{f} F_{cf} U - 2 \mathcal{A}^{c} \mathcal{A}^{f} U \mathcal{L}_n K_{cf} + 2  A_{*}^2 h^{cf} U \mathcal{L}_n K_{cf}\bigr)\\ +  U^{2} \mathcal{L}_nK_{cf}\big( \mathcal{A}^{c} \mathcal{A}^{f}-A_{*}^2 h^{cf}\big)+ A_{*}^2 \mathcal{F}_{2} U^2 - 2  \mathcal{A}^{c} A_{*} V_{c}U^{2} +  \mathcal{A}^{c} \mathcal{A}^{f} F_{cf} U^{2}\bigg]
\end{multline}
\begin{multline}
     L^{\prime}_{13}=  H_{13}\bigg[Z_{*}^2 \bigl( \mathcal{A}^{e} \mathcal{A}^{f} F_{ef} +  A_{*}^2 \mathcal{F}_{2} - 2  \mathcal{A}^{e} A_{*} V_{e} +\mathcal{A}^{e} \mathcal{A}^{f} \mathcal{L}_n K_{ef} -A_{*}^2 h^{ef}\mathcal{L}_n K_{ef}\bigr)+ A_{*}^2 \mathcal{F}_{2} U_{ef} U^{ef}\\+\mathcal{L}_n K_{ef}\big(\mathcal{A}^{e} \mathcal{A}^{f} U_{qs} U^{qs}-A_{*}^2 h^{ef} U_{qs} U^{qs}- 2  \mathcal{A}^{e} \mathcal{A}^{f} Y_{q} Y^{q}+ 2  A_{*}^2 h^{fe} Y_{q} Y^{q} \big) - 2\mathcal{A}^{e} A_{*} V_{e} U_{fq} U^{fq}\\ + \mathcal{A}^{e} \mathcal{A}^{f} F_{ef} U_{qs} U^{qs} - 2  A_{*}^2 \mathcal{F}_{2} Y_{e} Y^{e} + 4  \mathcal{A}^{e} A_{*} V_{e} Y_{f} Y^{f} - 2  \mathcal{A}^{e} \mathcal{A}^{f} F_{ef} Y_{q} Y^{q} \bigg]
\end{multline}
\begin{multline}
    L^{\prime}_{14}=  H_{14}\bigg[ Z_{*}\big( A_{*}^2 \mathcal{F}^{cd} U_{cd}-\mathcal{A}^{c} \mathcal{A}^{d} \mathcal{F}^{e}{}_{c} U_{de}+\mathcal{A}^{c} A_{*} V_{c}{}^{de} U_{de} + 2  \mathcal{A}^{c} A_{*} \mathcal{F}^{d}{}_{c} Y_{d}-A_{*}^2 U^{cd} \mathcal{L}_n K_{cd}\\+\mathcal{A}^{c} \mathcal{A}^{d}U_{c}{}^{e} \mathcal{L}_n K_{de})+  \mathcal{L}_n K_{de}\big( A_{*}^2 Y^{e} Y^{d} + \mathcal{A}^{c} A_{*} U^{de} Y_{c} - \mathcal{A}^{c} A_{*} U_{c}{}^{e} Y^{d}- \mathcal{A}^{c} \mathcal{A}^{d} Y_{c} Y^{e}\big)\\-\mathcal{A}^{c} A_{*} V^{def} U_{cd} U_{ef} -\mathcal{A}^{c}  \mathcal{A}^{d}\mathfrak{R}_{efdp} U_{c}{}^{e} U^{fp} -\mathcal{A}^{c} A_{*} \mathcal{F}^{de} U_{de} Y_{c} -   \mathcal{A}^{c}\mathcal{A}^{d} V_{d}{}^{ef} U_{ef} Y_{c}\\ + A_{*}^2 V_{c}{}^{de} U_{de}Y^{c} +  A_{*}^2 \mathcal{F}^{cd} Y_{c} Y_{d}+  \mathcal{A}^{c} A_{*} \mathfrak{R}_{decf} U^{ef} Y^{d} -\mathcal{A}^{c} A_{*} \mathcal{F}^{de} U_{cd} Y_{e}\\ - \mathcal{A}^{c} \mathcal{A}^{d}\mathcal{F}^{e}{}_{c} Y_{d} Y_{e} + \mathcal{A}^{c} \mathcal{A}^{d} V_{ce}{}^{f} U_{df} Y^{e} -\mathcal{A}^{c} \mathcal{A}^{d} V_{e}{}^{f}{}_{c} U_{df} Y^{e} - \mathcal{A}^{c} A_{*}V_{cde} Y^{d} Y^{e}\bigg]
\end{multline}
\begin{multline}
      L^{\prime}_{15}=  H_{15}\bigg[ Z_{*}^2 \big( \mathcal{A}^{c} \mathcal{A}^{d} \mathcal{F}_{cd} +  \mathcal{A}^{c} \mathcal{A}^{d} \mathcal{L}_nK_{cd})+Z_{*} \bigg\{ A_{*}^2 \mathcal{F}^{cd} U_{cd} - 2  \mathcal{A}^{c} A_{*} V_{c}{}^{df} U_{df} -   \mathcal{A}^{c} \mathcal{A}^{d} \mathcal{F}_{cd} U -   \mathcal{A}^{c} \mathcal{A}^{d}\mathfrak{R}_{cfdp} U^{fp}\\ +  \mathcal{A}^{c} A_{*} \mathcal{F}_{c}{}^{d} Y_{d} +  \mathcal{A}^{c} A_{*} \mathcal{F}^{d}{}_{c} Y_{d} + 2  \mathcal{A}^{c} \mathcal{A}^{d} V_{cfd} Y^{f} + \mathcal{L}_n K_{cd}\big( A_{*}^2 U^{cd}-\mathcal{A}^{c} \mathcal{A}^{d} U - 2  \mathcal{A}^{c} A_{*} Y^{d} \big)\bigg\}\\+ \mathcal{L}_n K_{cd}\big( 2  \mathcal{A}^{c} A_{*} U Y^{d}-A_{*}^2 U U^{cd}\big)-  A_{*}^2 \mathcal{F}^{cd} U_{cd} U + 2  \mathcal{A}^{c} A_{*} V_{c}{}^{df} U_{df} U +  \mathcal{A}^{c} \mathcal{A}^{d}\mathfrak{R}_{cpdq} U U^{pq}\\ -  \mathcal{A}^{c} A_{*} \mathcal{F}_{c}{}^{d} U Y_{d} -   \mathcal{A}^{c} A_{*} \mathcal{F}^{d}{}_{c} U Y_{d} - 2  \mathcal{A}^{c} \mathcal{A}^{d} V_{cfd} U Y^{f}\bigg]
\end{multline}
\begin{multline}
    L^{\prime}_{16}=  H_{16}\bigg[ Z_{*}^2 \bigl( \mathcal{A}^{c} \mathcal{A}^{d} \mathcal{F}_{cd} +  \mathcal{A}^{c} \mathcal{A}^{d} \mathcal{L}_n K_{cd}\bigr) + Z_{*} \big( \mathcal{A}^{c} A_{*} \mathcal{F}_{c}{}^{d} Y_{d} +  \mathcal{A}^{c} A_{*} \mathcal{F}^{d}{}_{c} Y_{d} + 2  \mathcal{A}^{c} \mathcal{A}^{d} V_{ced} Y^{e} - 2  \mathcal{A}^{c}A_{*} Y^{d} \mathcal{L}_n K_{cd}\big)\\+ \mathcal{L}_n K_{cd}\big( A_{*}^2 Y^{c} Y^{d} - \mathcal{A}^{c} \mathcal{A}^{d} Y_{e} Y^{e}+ 2  \mathcal{A}^{c} A_{*} U_{e}{}^{d} Y^{e} - A_{*}^2 U_{e}{}^{c} U^{ed}\big)-  A_{*}^2 \mathcal{F}^{cd} U_{c}{}^{e} U_{de} + 2  \mathcal{A}^{c} A_{*} V_{c}{}^{de}  U_{d}{}^{f} U_{ef} \\+  \mathcal{A}^{c} \mathcal{A}^{d}\mathfrak{R}_{cfdp} U_{e}{}^{p} U^{ef} + A_{*}^2 \mathcal{F}^{cd} Y_{c} Y_{d} -   \mathcal{A}^{c} A_{*} \mathcal{F}_{c}{}^{d} U_{de} Y^{e} -   \mathcal{A}^{c} A_{*} \mathcal{F}^{d}{}_{c} U_{de} Y^{e} - 2  \mathcal{A}^{c} \mathcal{A}^{d} V_{c}{}^{f}{}_{d} U_{ef} Y^{e}\\ - 2  \mathcal{A}^{c} A_{*} V_{cde} Y^{d} Y^{e}-   \mathcal{A}^{c} \mathcal{A}^{d} \mathcal{F}_{cd} Y_{e} Y^{e} -   \mathcal{A}^{c} \mathcal{A}^{d}\mathfrak{R}_{cedf} Y^{e} Y^{f}\bigg]
\end{multline}

\bibliographystyle{unsrtnat}
\bibliography{refs}

\end{document}